\documentclass[11pt]{article}
\usepackage[utf8]{inputenc}
\usepackage[margin=1.1in]{geometry}
\usepackage{amsmath}
\usepackage{amssymb}
\usepackage{amsthm}
\usepackage{bbm}

\usepackage{multicol}
\usepackage{multirow}
\usepackage{booktabs}
\usepackage{graphicx}
\usepackage{float}
\usepackage{bm}
\usepackage{multirow}
\usepackage{xargs}
\usepackage{listings}
\usepackage{xcolor}
\usepackage{tabularx}
\usepackage{algorithm}
\usepackage{algcompatible}
\usepackage{rotating}
\usepackage{comment}
\usepackage{textcomp}

\usepackage{bbm}
\usepackage{comment}
\usepackage{natbib}
\usepackage[colorlinks=true, allcolors=blue]{hyperref}

\theoremstyle{definition}

\newcommand{\footremember}[2]{
    \footnote{#2}
    \newcounter{#1}
    \setcounter{#1}{\value{footnote}}
}
\newcommand{\footrecall}[1]{
    \footnotemark[\value{#1}]
} 

\providecommand{\keywords}[1]
{
  \small	
  \textbf{\textit{Keywords---}} #1
}

\title{Bayesian Prognostic Covariate Adjustment With Additive Mixture Priors
\author{Alyssa M. Vanderbeek\footnote{Corresponding author} \footremember{Unlearn.AI}{Unlearn.AI, Inc., San Francisco, CA, USA}, \and Arman Sabbaghi\footrecall{Unlearn.AI}, \and Jon R. Walsh\footrecall{Unlearn.AI}, \and Charles K. Fisher\footrecall{Unlearn.AI}}}
\date{\today}

\begin{document}

\maketitle
\begin{abstract}
Effective and rapid decision-making from randomized controlled trials (RCTs) requires unbiased and precise treatment effect inferences. Two strategies to address this requirement are to adjust for covariates that are highly correlated with the outcome, and to leverage historical control information via Bayes' theorem. We propose a new Bayesian prognostic covariate adjustment methodology, Bayesian PROCOVA, that combines these two strategies. Covariate adjustment in Bayesian PROCOVA is based on generative artificial intelligence (AI) algorithms that construct a digital twin generator (DTG) for RCT participants. The DTG is trained on historical control data and yields a digital twin (DT) probability distribution for the outcome on the control treatment for each RCT participant's. This distribution is the digital twin (DT) and the  expectation, referred to as the prognostic score, defines the covariate for adjustment. Historical control information is leveraged via an additive mixture prior with two components: an informative prior specified based on historical control data, and a weakly informative prior. The mixture weight determines how often posterior inferences are drawn from the informative versus weakly informative component. This weight has a prior distribution as well, and so the entire additive mixture prior is completely pre-specifiable without involving any RCT data. We establish an efficient Gibbs algorithm for sampling from the posterior distribution, and derive closed-form expressions for the conditional posterior mean and variance of the treatment effect parameter. Via simulations, we demonstrate Bayesian PROCOVA to have bias control and variance reduction compared to frequentist PROCOVA. These gains can be translated to smaller RCTs. We also describe how hyperparameters can be specified to target expected operating characteristics, conditional on the quality of the DTG. Ultimately, Bayesian PROCOVA can yield informative treatment effect inferences with fewer control participants, thereby accelerating effective decision-making from RCTs. 
\end{abstract}

\section{Improved Decision-Making With Randomized Controlled Trials}
\label{sec:introduction}

Randomized controlled trials (RCTs) are increasingly faulted for failing to enable effective and rapid decision-making \citep{frieden_2017, fisher_2023, subbiah_2023}. Key stakeholders in drug development are pushing for innovations in RCTs to address concerns of premature or incorrect decision-making that could lead to the abandonment of truly efficacious medical treatments, in addition to many other concerns and issues described by \citet{fogel_2018}. The COVID-19 pandemic further established the urgency for innovations to improve and accelerate decision-making from RCTs \citep{fernando_2022}. Statistical inferences for treatment effects in RCTs underlie decision-making in drug development, and so an imperative to address the faults of RCTs is to develop innovative statistical methods that yield unbiased treatment effect inferences with reduced uncertainty.

Several strategies exist to obtain unbiased and precise treatment effect inferences for improved decision-making from RCTs. Two effective and general strategies are to adjust for participant covariates (e.g., baseline information collected at the start of an RCT), and to augment the RCT information with historical control information via Bayes' theorem. 

Regulatory agencies recognize covariate adjustment as a valid statistical method for unbiased and precise treatment effect inference, with the caveat that the adjustment should incorporate an appropriately small number of covariates \citep{ema_adjusting_2015, food_and_drug_administration_adjusting_2023}. An innovative approach for covariate adjustment is to use generative artificial intelligence (AI) algorithms, trained on historical control data, to yield a function of covariates that is optimized in terms of its correlation with the control outcomes. \citet{schuler_2022} describe their statistical methodology of prognostic covariate adjustment (PROCOVA\texttrademark) that implements this approach for RCTs. The generative AI algorithm that they consider yields a digital twin generator (DTG) whose inputs are a participant's (potentially high-dimensional) covariate vector and whose output is a digital twin (DT) probability distribution for the participant's control outcome. Under PROCOVA, the mean of the DT distribution is calculated for each RCT participant and defines the single, optimized covariate that is used for adjustment in the analysis of the RCT. This covariate is referred to as the prognostic score. The European Medicines Agency (EMA) qualified PROCOVA as ``an acceptable statistical approach for primary analysis'' of Phase 2/3 RCTs with continuous endpoints \citep{ema_procova_2022}. 

The second strategy, involving Bayesian inference for the treatment effect, is increasing in consideration for modern RCTs, although fewer practitioners may be as familiar with the new Bayesian methods as with established frequentist methods \citep{muehlemann_2023}. The Food and Drug Administration (FDA)'s Center for Devices and Radiological Health (CDRH) and Center for Biologics Evaluation and Research (CBER) published guidance, with informative explanatory materials, on Bayesian inference for medical device trials in 2010 \citep{food_and_drug_administration_bayesian_2010}. The FDA Center for Drug Evaluation and Research (CDER) and CBER have yet to publish guidance documents on Bayesian methods for applications beyond medical device trials (which are considered the domain of CDRH), but \citet{ionan_2023} and \citet{travis_2023} discuss relevant considerations for the use of Bayesian methods. The Bayesian methods that were recently developed by \citet{egidi_2022}, \citet{zhao_2023}, and \citet{yang_2023}, all three of which utilize additive mixture priors, reflect these regulatory considerations. Most notably, the approval of Pfizer's COVID-19 vaccine involved Bayesian analyses \citep{polack_2020}, and highlighted the utility of Bayesian inference over frequentist methods for effective and rapid decision-making.

These two strategies have yet to be combined to advance decision-making from RCTs. \citet{walsh_2020} proposed a Bayesian version of PROCOVA, but their prior distribution is not particularly justifiable or interpretable with respect to the regulatory considerations and examples of Bayesian analyses provided in \citep{ionan_2023, travis_2023}. The Bayesian methods of \citet{egidi_2022}, \citet{zhao_2023}, and \citet{yang_2023} cannot be completely specified prior to the commencement of the RCT because the essential ingredient of the ``weight'' parameter in their additive mixture prior is specified based on RCT data. In particular, \citet[p.~494, 498]{egidi_2022} and \citet{zhao_2023} define the weight as a $p$-value based on the RCT data, whereas \citet{yang_2023} define the weight as a likelihood ratio test statistic involving the RCT data. Their specifications in this regard lead to twice the use of the RCT data, which technically constitutes an improper application of Bayes' theorem and complicates the regulatory approval process and interpretations of uncertainty quantifications from the Bayesian analysis. Limitations also exist in their scopes of application. For example, \citet{zhao_2023} and \citet{yang_2023} consider solely binary endpoints, and do not incorporate covariate adjustment as in a regression model. The suggested approach for covariate adjustment in the method of \citet{yang_2023} is propensity score matching, which may not be acceptable or desirable in practice. A gap remains in defining a completely pre-specificable, fully Bayesian covariate adjustment methodology that can incorporate predictors from generative AI algorithms for the analysis of continuous outcomes.

We propose a new Bayesian methodology to perform covariate adjustment via the prognostic score and to leverage historical control information (consisting both of prognostic scores and control outcomes) in the analysis of an RCT. We refer to this method as Bayesian PROCOVA, as it constitutes a fully Bayesian extension of PROCOVA. Following the work of \citet{egidi_2022}, \citet{yang_2023}, and \citet{zhao_2023}, we encode historical control information in Bayesian PROCOVA via one component of an additive mixture prior, and set the second component to be weakly informative. This prior specification results in a posterior distribution that dynamically borrows information from historical control data, thereby effectively augmenting the information in the RCT. When historical control and RCT data are consistent, Bayesian PROCOVA puts significant weight on the information from the historical control data and consequently increases the precision of treatment effect inferences. When historical control and RCT data are discrepant, Bayesian PROCOVA discounts the historical control data and yields inferences similar to PROCOVA in terms of controlling bias and precision in treatment effect inferences. Finally, a prior distribution is specified for the mixture weight parameter in Bayesian PROCOVA so as to make the entire prior completely pre-specifiable and interpretable before the commencement of the RCT, and to yield a fully Bayesian analysis. 

Thus Bayesian PROCOVA effectively addresses the limitations of the Bayesian methods of \citet{egidi_2022}, \citet{zhao_2023}, and \citet{yang_2023}, as its additive mixture prior distribution is independent of any information from the RCT and still enables interpretable covariate adjustment and dynamic information borrowing for continuous outcomes. In addition, Bayesian PROCOVA goes beyond existing Bayesian methodologies by including the optimized prognostic score in the analysis of the RCT. It is particularly advantageous for improving the quality of treatment effect inferences, and hence decision-making, from small RCTs. 

Section \ref{sec:background} provides notations, assumptions, and background materials for Bayesian PROCOVA. The methodology is described in Section \ref{sec:Bayesian_PROCOVA}, including closed-form formulae for the posterior mean and variance of the treatment effect parameter conditional on the mixture weight (Section \ref{sec:posterior_distributions)}, and an outline of a Gibbs algorithm \citep{geman_1984} for calculating the joint posterior distribution of all the parameters in Section \ref{sec:gibbs_sampler}. In the Appendix, we additionally describe a quantitatively robust way of tuning the informative prior to control for a maximum Type I error rate. Bias control and variance reduction of the treatment effect estimator from Bayesian PROCOVA compared to PROCOVA is shown via extensive simulation studies in Section \ref{sec:simulation_experiments}. In these studies, we demonstrate how the properties of Bayesian PROCOVA change in cases of discrepancies between the historical control and RCT data due to domain shifts, and due to changes in correlations between the prognostic scores and the control outcomes. As we conclude in Section \ref{sec:concluding_remarks}, Bayesian PROCOVA can improve the quality of treatment effect inferences compared to frequentist methods, and thereby can advance decision-making from smaller and faster RCTs.

\section{Background}
\label{sec:background}

\subsection{Notations and Assumptions}
\label{sec:notations_assumptions}

Bayesian PROCOVA is formulated under the Neyman-Rubin Causal Model \citep{neyman_1923, rubin_1974, holland_1986}. To describe this methodology, we first define the experimental units, covariates, treatments, potential outcomes, and causal estimands under consideration for the RCT and historical control data. We also provide the assumptions that we invoke on these elements in order to facilitate causal inferences via Bayesian PROCOVA.

Experimental units are participants in the RCT at a particular time-point \citep[p.~4]{imbens_rubin_2015}. Each RCT participant $i = 1, \ldots, N$ has a vector of covariates $x_i \in \mathbb{R}^L$ that are either measured prior to treatment assignment, or measured afterwards but are known to be unaffected by treatment \citep[p.~15--16]{imbens_rubin_2015}. The treatment indicator for participant $i$ is $w_i \in \{0, 1\}$ for $0$ (control) and $1$ (active treatment). For each participant $i$ and treatment option $w$, we define potential outcome $Y_i(w)$ as their endpoint value that would be observed at a specified time-point after treatment assignment. We invoke the Stable Unit-Treatment Value Assumption \citep[SUTVA,][p.~9--13]{imbens_rubin_2015} in this definition, so that the pair of potential outcomes $Y_i = \left ( Y_i(0), Y_i(1) \right )^{\mathsf{T}}$ of each participant is well-defined.

Causal estimands are defined as comparisons of potential outcomes for a set of experimental units \citep[p.~18--19]{imbens_rubin_2015}. The experimental units could correspond to the finite-population of participants in the RCT, or to the conceptual ``super-population'' of all potential RCT participants. An estimand defined on the former set of participants is a finite-population estimand, and an estimand defined on the latter is a super-population estimand. For example, the quantity $\bar{Y}(1) - \bar{Y}(0) = N^{-1} \sum_{i=1}^N \left \{ Y_i(1) - Y_i(0) \right \}$ is the finite-population average treatment effect. To define the super-population average treatment effect, let $\mu \left ( w \right ) = \int_{-\infty}^{\infty} y dF_w(y)$ denote the expected value of the potential outcomes under treatment $w \in \{0, 1\}$ for the super-population of participants as defined by the cumulative distribution function $F_w: \mathbb{R} \rightarrow (0,1)$. Then the super-population average treatment effect is  $\mu(1) - \mu(0)$. We focus on inferring $\mu(1) - \mu(0)$ via Bayesian PROCOVA in Section \ref{sec:Bayesian_PROCOVA}, and summarize in Section \ref{sec:bayesian_statistics} how Bayesian inference can be conducted for $\bar{Y}(1) - \bar{Y}(0)$.

In addition to RCT data, Bayesian PROCOVA incorporates information from a historical control dataset, i.e., a dataset independent of the RCT in which all participants are given control. These data are used to specify one component of the mixture prior distribution for the model parameters in Bayesian PROCOVA. For the historical control data we denote the sample size by $N_H$, the covariate vector for participant $i = 1, \ldots, N_H$ by $x_{i,H} \in \mathbb{R}^L$, and their outcome by $y_{i,H}$.

Causal inference under the Neyman-Rubin Causal Model is a missing data problem, because at most one potential outcome can be observed for any participant \citep{holland_1986}. The treatment assignment mechanism, i.e., the probability mass function $p( w_1, \ldots, w_N \mid Y_1, \ldots, Y_N, x_1, \ldots, x_N)$, is critical for obtaining valid causal inferences. Similar to other types of missing data problems, it is essential to consider the treatment assignment mechanism so as to specify the likelihood function for Bayesian PROCOVA \citep[p.~39, 152--156]{imbens_rubin_2015}. We assume that the treatment assignment mechanism is probabilistic, individualistic, and unconfounded \citep[p.~37--39]{imbens_rubin_2015}. These three assumptions are generally valid for traditional RCTs, and correspond to a strongly ignorable missing data mechanism \citep[p.~39]{rosenbaum_1983, imbens_rubin_2015}. They also enable the ``automated'' specification of the likelihood function for Bayesian inference on causal estimands, in terms of the observed outcomes $y_i = w_iY_i(1) + \left ( 1 - w_i \right )Y_i(0)$ in the RCT \citep[p.~43--44]{rubin_1978}.

\subsection{Bayesian Inference and Linear Regression}
\label{sec:bayesian_statistics}

Bayesian inference refers to the fitting of a statistical model to data so as to obtain a probability distribution on the unknown model parameters $\theta$ \citep[p.~1]{gelman_2013}. Under this paradigm, all uncertainties and information are encoded via probability distributions, and all inferences and conclusions are obtained via the laws of probability theory. The essential characteristic of Bayesian inference is its direct and explicit use of probability, specifically, via the prior and posterior probability distributions for $\theta$, for quantifying uncertainty and information for all unknown parameters. This characteristic of Bayesian inference distinguishes it from frequentist inference, in which probability distributions are generally not specified for $\theta$. 

The necessary elements for Bayesian inference are the prior distribution $p \left ( \theta \right )$ for $\theta$ and the likelihood function for $\theta$. The prior encodes information about $\theta$ that is contained in historical data, and is expected to augment the information from the RCT. We denote the likelihood function by $L \left ( \theta \mid y, w, X \right )$, where $y = \left ( y_1, \ldots, y_N \right )^{\mathsf{T}}$ is the vector of the participants' observed outcomes, $w = \left ( w_1, \ldots, w_N \right )^{\mathsf{T}}$ is the vector of their treatment assignments, and $X = \begin{pmatrix} x_1^{\mathsf{T}} \\ \vdots \\ x_N^{\mathsf{T}} \end{pmatrix}$ is the matrix of their covariates. The likelihood function encodes information about $\theta$ from the RCT, and is obtained from the sampling distribution of the data as specified by the generative statistical model underlying the analysis \citep[p.~6--8]{gelman_2013}. The prior and likelihood function are combined via Bayes' theorem to calculate the posterior distribution $p \left ( \theta \mid y, w, X \right )$ of $\theta$ conditional on the data. In practice, the posterior distribution is calculated as a proportional quantity, without a normalization constant, according to $p \left ( \theta \mid y, w, X \right ) \propto p \left ( \theta \right ) L \left ( \theta \mid y, w, X \right )$. 

The posterior distribution $p \left ( \theta \mid y, w, X \right )$ encodes all information about $\theta$ that is contained in the prior and data. All inferences on causal estimands can thus be obtained from this distribution. To illustrate, consider the finite-population average treatment effect $\bar{Y}(1) - \bar{Y}(0)$. Bayesian inference for this estimand requires the calculation of its posterior distribution. Following the framework and reasoning for Bayesian causal inference employed by \citet[p.~43--45]{rubin_1978} and \citet[p.~153--155]{imbens_rubin_2015}, this posterior distribution is calculated by repeatedly drawing from $p \left ( \theta \mid y, w, X \right )$, using the draws to impute the missing potential outcomes $y_i^{\mathrm{mis}} = \left ( 1 - w_i \right )Y_i(1) + w_i Y_i(0)$, calculating the estimand for each such imputation according to $N^{-1}\sum_{i=1}^N \left \{ \left ( 2w_i-1 \right ) \left ( y_i - y_i^{\mathrm{mis}} \right ) \right \}$, and concatenating all such calculated estimands. More formally, the posterior $p \left ( \bar{Y}(1) - \bar{Y}(0) \mid y, w, X \right )$ is calculated according to the integration
\[
\int p \left ( \bar{Y}(1) - \bar{Y}(0) \mid y^{\mathrm{mis}}, \theta, y, w, X \right ) p \left ( y^{\mathrm{mis}} \mid \theta, y, w, X \right ) p \left ( \theta \mid y, w, X \right ) dy^{\mathrm{mis}} d\theta,
\]
where $y^{\mathrm{mis}} = \left ( y_1^{\mathrm{mis}}, \ldots, y_N^{\mathrm{mis}} \right )^{\mathsf{T}}$. Given this posterior distribution, point estimates of the causal estimand can be obtained via its mean, median, mode(s), and other functionals of the posterior distribution. Interval estimates can be obtained by computing quantiles of the posterior distribution. 

Bayesian inferences for the super-population average treatment effect $\mu(1) - \mu(0)$ are obtained from the posterior distribution for a specified parameter in the data generating mechanism. The mechanism that we consider is the linear regression model
\begin{equation}
\label{eq:data_generating_model}
Y_i(w) = v_i^{\mathsf{T}}\beta + \epsilon_i(w),
\end{equation}
where $v_i \in \mathbb{R}^K$ is the vector of predictors for participant $i = 1, \ldots, N$ that are defined as functions of $w_i$ and $x_i$, $\beta = \left ( \beta_0, \ldots, \beta_{K-1} \right )^{\mathsf{T}}$ is the vector of regression coefficients, and the $\epsilon_i(w)$ are independent random error terms distributed according to $\left [ \epsilon_i(w) \mid X, \beta, \sigma^2 \right ]\sim \mathrm{N} \left ( 0, \sigma^2 \right )$ with variance parameter $\sigma^2 > 0$. The first two entries in each $v_i$ are $v_{i,1} = 1$ and $v_{i,2} = w_i$, and the $\beta_1$ entry in $\beta$ corresponds to the super-population average treatment effect when there are no interactions between treatment and covariates in $v_i$. As the observed outcomes are functions of the potential outcomes and treatment indicators, the mechanism in equation (\ref{eq:data_generating_model}) motivates the linear regression model 
\begin{equation}
\label{eq:linear_regression_model}
y_i = v_i^{\mathsf{T}} \beta + \epsilon_i,
\end{equation}
for the observed outcomes, where the $\left [ \epsilon_i \mid X, \beta, \sigma^2 \right ]\sim \mathrm{N} \left ( 0, \sigma^2 \right )$ independently as before. Bayesian inferences are performed on $\beta_1$ (and other parameters) by extending model (\ref{eq:linear_regression_model}) with a prior distribution on $\beta, \sigma^2$ and calculating the posterior distribution according to Bayes' theorem. The unconfoundedness assumption automates the derivation of the likelihood function as
\begin{equation}
\label{eq:regression_likelihood}
L \left ( \beta, \sigma^2 \mid y, w, X \right ) = \left ( \sigma^2 \right )^{-N/2} \mathrm{exp} \left \{ -\frac{1}{2\sigma^2} \left ( y - V \beta \right )^{\mathsf{T}} \left ( y - V \beta \right ) \right \},
\end{equation}
where $V = \begin{pmatrix} v_1^{\mathsf{T}} \\ \vdots \\ v_N^{\mathsf{T}} \end{pmatrix}$. The posterior distribution is then calculated according to
\begin{equation}
\label{eq:regression_posterior}
p \left ( \beta, \sigma^2 \mid y, w, X \right ) \propto p \left ( \beta, \sigma^2 \right ) \left ( \sigma^2 \right )^{-N/2} \mathrm{exp} \left \{ -\frac{1}{2\sigma^2} \left ( y - V \beta \right )^{\mathsf{T}} \left ( y - V \beta \right ) \right \}.
\end{equation}
The standard non-informative (and improper) prior for the model parameters is $p \left ( \beta, \sigma^2 \right ) \propto \left ( \sigma^{2} \right )^{-1}$, and this corresponds to independent, flat priors on $\beta$ and $\mathrm{log} \left ( \sigma^2 \right )$. Inferences from PROCOVA (excluding the heteroskedastic-consistent \citep[HC,][]{white_1980, romano_2017}, or robust, standard errors \citep[p.~333]{schuler_2022}) are equivalent to posterior inferences from the corresponding Bayesian linear regression with this prior \citep[p.~355--356]{gelman_2013}. \citet[p.~353--380]{gelman_2013} provide additional computational techniques and inferential procedures for Bayesian linear regression.

\subsection{Prognostic Covariate Adjustment}
\label{sec:procova}

The merits of regression models for causal inference are long-established in the literature(\citet{yule_1899}, \citep[p.~247--250]{freedman_1999}) and are being recognized by regulatory agencies. The FDA's guidance on covariate adjustment states that ``Covariate adjustment leads to efficiency gains when the covariates are prognostic for the outcome of interest in the trial. Therefore, FDA recommends that sponsors adjust for covariates that are anticipated to be most strongly associated with the outcome of interest.'' \citep[p.~3]{food_and_drug_administration_adjusting_2023}. Similarly, the EMA's guideline document states that ``Variables known a priori to be strongly, or at least moderately, associated with the primary outcome and/or variables for which there is a strong clinical rationale for such an association should also be considered as covariates in the primary analysis.'' \citep[p.~3]{ema_adjusting_2015}. 

Both agencies also issued provisos that the number of covariates for adjustment should be kept at an appropriate minimum and should be highly correlated with the outcome. Specifically, the FDA states that ``The statistical properties of covariate adjustment are best understood when the number of covariates adjusted for in the study is small relative to the sample size \citep{tsiatis_2008}.'' \citep[p.~4]{food_and_drug_administration_adjusting_2023}. The EMA states more directly that ``Only a few covariates should be included in a primary analysis. Although larger data sets may support more covariates than smaller ones, justification for including each of the covariates should be provided.'' and ``The primary model should not include treatment by covariate interactions. If substantial interactions are expected a priori, the trial should be designed to allow separate estimates of the treatment effects in specific subgroups.'' \citep[p.~3--4]{ema_adjusting_2015}.

The use of generative AI algorithms in PROCOVA directly addresses these fundamental regulatory considerations for covariate adjustment in RCTs. The generative AI algorithm is trained on historical control data, and the sole inputs for generating DTs are the participants' baseline covariates. These two aspects ensure that bias cannot result from the use of the DTG outputs, and that their use for covariate adjustment corresponds to regulatory guidance on pre-specifying all aspects of trial design and analysis prior to the commencement of the RCT. The DTG outputs for an RCT participant are forecasts for their control outcomes at future time-points after treatment assignment. The forecasts at one time-point correspond to the participant's DT probability distribution, and summaries of the DT distribution are used for adjustment. These summaries are themselves covariates, as they are transformations of baseline covariates. We denote the DT distribution at a specified time-point for participant $i$ by the cumulative distribution function $G_i: \mathbb{R} \rightarrow (0,1)$. By virtue of the training process for the DTG, the prognostic score $m_i = \int_{-\infty}^{\infty} ydG_i(y)$ is an optimized transformation of a participant's covariates in terms of its absolute correlation with the control outcome. This feature is advantageous for, and follows regulatory guidance on, covariate adjustment because it summarizes the information in a high-dimensional covariate vector into a scalar variable for adjustment that is highly correlated with the outcome. The PROCOVA methodology of \citet{schuler_2022} leverages the prognostic score as the essential predictor in a linear regression analysis of a RCT, i.e., it sets $v_i = \left ( 1, w_i, m_i \right )^{\mathsf{T}}$ as in
\begin{equation}
\label{eq:procova}
y_i = \beta_0 + \beta_1 w_i + \beta_2 m_i + \epsilon_i.
\end{equation}
Inferences and tests for the treatment effect are performed with respect to $\beta_1$ in PROCOVA. Following regulatory guidance on uncertainty quantification for frequentist covariate adjustment \citep[p.~4--5]{food_and_drug_administration_adjusting_2023}, PROCOVA utilizes HC standard errors for inferences on $\beta_1$ \citep{schuler_2022}.

PROCOVA effectively leverages aspects of historical control data via covariate adjustment using AI-generated prognostic scores to improve the precision of unbiased treatment effect inferences. Further gains in precision beyond those from PROCOVA can be realized by combining covariate adjustment using the prognostic score with prior information from historical control data on the $\beta_0, \beta_2$, and $\sigma^2$ parameters in the PROCOVA model (\ref{eq:procova}). This combination is the defining feature of Bayesian PROCOVA, which we proceed to describe.

\section{Bayesian Prognostic Covariate Adjustment}
\label{sec:Bayesian_PROCOVA}

\subsection{Overview}
\label{sec:Bayeian_PROCOVA_overview}

Bayesian PROCOVA is a Bayesian extension of the PROCOVA model (\ref{eq:procova}) with an additive mixture prior for the parameters $\beta$ and $\sigma^2$ that is defined as the weighted sum of two probability density functions. The ``informative prior component'' $p_I \left ( \beta, \sigma^2 \right )$ is specified based on prognostic score and outcome information from historical control data. The ``flat prior component'' $p_F \left ( \beta, \sigma^2 \right )$ is specified independently of any data and serves as a proper, weakly informative prior. A mixture weight parameter $\boldsymbol{\omega} \in (0, 1)$ is given its own prior $p \left ( \boldsymbol{\omega} \right )$ that does not involve any RCT data. Thus, the joint prior for all unknown parameters in Bayesian PROCOVA is
\begin{equation}
\label{eq:additive_mixture_prior}
p \left ( \beta, \sigma^2, \boldsymbol{\omega} \right ) = \boldsymbol{\omega} p_I \left ( \beta, \sigma^2 \right ) p \left ( \boldsymbol{\omega} \right ) + \left ( 1 - \boldsymbol{\omega} \right ) p_F \left ( \beta, \sigma^2 \right ) p \left ( \boldsymbol{\omega} \right ).
\end{equation}
The additive mixture prior for Bayesian PROCOVA is specified so as to yield dynamic information borrowing \citep{yang_2023}. More formally, in the calculation of the posterior distribution for the regression parameters under Bayesian PROCOVA, the weight that is placed on the historical control information is effectively a function of the consistency between the historical control and RCT data. If the historical control and RCT data are consistent, then significant weight is placed on the information encoded in $p_I \left ( \beta, \sigma^2 \right )$ when calculating the posterior distribution, and the precision for $\beta_1$ consequently increases. Alternatively, if the historical control and RCT data are discrepant, then the information from $p_I \left ( \beta, \sigma^2 \right )$ is discounted when calculating the posterior distribution, and instead the weak information from $p_F \left ( \beta, \sigma^2 \right )$ is more highly weighted. 

\citet[p.~4--5]{travis_2023} note this attractive property of additive mixture priors in moving the posterior distribution towards the most compatible component rather than just towards historical information. As we demonstrate via extensive simulation studies in Section \ref{sec:simulation_experiments}, the combination of dynamic information borrowing with the tuning of the hyperparameters in $p_I \left ( \beta, \sigma^2 \right )$ helps to balance the two objectives of controlling the bias and increasing the precision of treatment effect inferences based on the level of consistency between historical control and RCT data in Bayesian PROCOVA.

\subsection{Likelihood and Prior Functional Forms}
\label{sec:Bayesian_PROCOVA_model}

The likelihood function for Bayesian PROCOVA is specified by modifying the PROCOVA model (\ref{eq:procova}) to imbue the intercept $\beta_0$ with an interpretation involving the prognostic scores (\citet[p.~4]{walsh_2020}). The essential sampling distribution for Bayesian PROCOVA is
\begin{equation}
\label{eq:bayesian_procova_model}
y_i = \beta_0 + \beta_1 w_i + \beta_2 \left ( m_i - \bar{m} \right ) + \bar{m} + \epsilon_i,
\end{equation}
so that the covariate adjustment for participant $i$ is their centered prognostic score $m_i - \bar{m}$, where $\bar{m} = N^{-1}\sum_{i=1}^N m_i$. This is equivalent to the regression model in which the observed outcomes are transformed according to $y_i^{(c)} = y_i - \bar{m}$ and the predictor vector is $v_i = \left ( 1, w_i, m_i - \bar{m} \right )^{\mathsf{T}}$. Here we consider the case of adjusting solely for the centered prognostic scores and with no interactions, but the theory can be expanded in a straightforward manner to include additional covariates and interactions. 

Parameter $\beta_0$ is interpreted as the bias of the average of the prognostic scores in predicting the endpoint of a control participant, i.e., $\beta_0 = \mathbb{E} \left ( y_i - \bar{m} \mid w_i = 0, \beta, \sigma^2, \boldsymbol{\omega} \right )$. For this interpretation we assume the $m_i$ are independent and identically distributed, with finite mean, and that their probability distribution does not depend on the model parameters. The likelihood function (\ref{eq:regression_likelihood}) corresponding to the model in equation (\ref{eq:bayesian_procova_model}) (excluding proportionality constants) is
\begin{equation}
\label{eq:bayesian_procova_likelihood}
L \left ( \beta, \sigma^2, \boldsymbol{\omega} \mid y, w, X \right) = \left (\sigma^2 \right )^{-N/2} \mathrm{exp} \left \{ -\frac{1}{2\sigma^2} \left ( y^{(c)} - V\beta \right )^{\mathsf{T}} \left ( y ^{(c)} - V \beta \right ) \right \}
\end{equation}
where $y^{(c)} = \left ( y_1^{(c)}, \ldots, y_N^{(c)} \right )^{\mathsf{T}}$ and $V = \begin{pmatrix} 1 & w_1 & \left ( m_1 - \bar{m} \right ) \\ \vdots & \vdots & \vdots \\ 1 & w_N & \left ( m_N - \bar{m} \right ) \end{pmatrix}$. 

To formally specify the prior $p \left ( \beta, \sigma^2, \boldsymbol{\omega} \right )$, we extend notation from the RCT to the historical control data. Let $m_{i,H} \in \mathbb{R}$ denote the prognostic score for participant $i$ in the historical control data, $\bar{m}_H = N_H^{-1} \sum_{i=1}^{N_H} m_{i,H}$ be the average of the historical prognostic scores, $y_{i,H}^{(c)} = y_{i,H} - \bar{m}_H$ be the historical control outcomes centered by the average of the historical prognostic scores, and $v_{i,H} = (1, m_{1,H} - \bar{m}_H )^{\mathsf{T}}$.

First, we specify $p_I \left ( \beta, \sigma^2 \right )$ according to $\left [ \beta \mid \sigma^2 \right ] \sim \mathrm{N} \left ( \begin{pmatrix} \widehat{\beta_{0,H}} \\ 0 \\ \widehat{\beta_{2, H}} \end{pmatrix}, \sigma^2 \boldsymbol{K} \right )$, where  $\sigma^2 \sim (N_H-2)s_H^2/\chi_{N_H-2}^2$ and $\boldsymbol{K} = \text{diag}(K_{0,H}, K_{1,H}, K_{2,H})$  is a $3 \times 3$ diagonal matrix of positive constants. Specifically, $\begin{pmatrix} \widehat{\beta_{0,H}} \\ \widehat{\beta_{2,H}} \end{pmatrix} = \left ( V_H^{\mathsf{T}}V_H \right )^{-1} V_H^{\mathsf{T}}y_H^{(c)}$ and $s_H^2 = \left ( N_H-2 \right )^{-1} \sum_{i=1}^N \left \{ y_{i,H}^{(c)} - \widehat{\beta_{0,H}} - \widehat{\beta_{2,H}} \left ( m_{i,H} - \bar{m}_H \right ) \right \}^2$ is the point estimate of $\sigma^2$ from the historical control data. We describe $\boldsymbol{K}$ in greater detail in Section \ref{sec:hyperparameters}.

Under this prior, the marginal distribution of $y$ is an $N$-dimensional Multivariate $t$ distribution with $N_H-2$ degrees of freedom, center equal to $V \begin{pmatrix} \widehat{\beta_{0,H}} \\ 0 \\ \widehat{\beta_{2,H}} \end{pmatrix} + \bar{m}\mathbf{1}$ (where $\mathbf{1}$ is the $N \times 1$ vector whose entries are all $1$), and scale matrix $s_H^2 \left ( I_{N \times N} + V \boldsymbol{K} V^{\mathsf{T}} \right )$.

Next, we specify $p_F \left ( \beta, \sigma^2 \right )$ according to $\left [ \beta \mid \sigma^2 \right ] \sim \mathrm{N} \left ( \mathbf{0}, \sigma^2 k I_{3 \times 3} \right )$ and $\sigma^2 \sim \nu_0\sigma_0^2/\chi_{\nu_0}^2$. Hyperparameter $k$ governs the prior variances of $\beta_0, \beta_1$, and $\beta_2$, and its selection should correspond to a large value with respect to the scale of the endpoint. The value of $\sigma_0^2$ is interpreted as a prior point estimate of $\sigma^2$ with $\nu_0$ degrees of freedom. As $\nu_0 \rightarrow 0$ for a fixed $\sigma_0^2$, the prior converges to the standard non-informative prior. We could take this limiting case and set the prior for $\sigma^2$ in the flat component as the standard non-informative prior. However, as for the informative component, in Bayesian PROCOVA we specify the flat component so that it is fully generative and a proper probability distribution. 

Under this prior, the marginal distribution of $y$ is an $N$-dimensional Multivariate $t$ distribution with $\nu_0$ degrees of freedom, center at $\bar{m}\mathbf{1}$, and scale matrix equal to $\sigma_0^2 \left ( I_{N \times N} + kVV^{\mathsf{T}} \right )$.

Finally, we specify the prior on the mixture weight $\boldsymbol{\omega}$ such that it does not depend on any information from the RCT. A flexible and established class of prior distributions consists of the $\mathrm{Beta} \left ( \alpha_1, \alpha_2 \right )$ distributions, with probability density function $p \left ( \boldsymbol{\omega} \right ) = \Gamma \left ( \alpha_1 + \alpha_2 \right ) \Gamma \left ( \alpha_1 \right )^{-1} \Gamma \left ( \alpha_2 \right )^{-1} \boldsymbol{\omega}^{\alpha_1 - 1} \left ( 1 - \boldsymbol{\omega} \right )^{\alpha_2 - 1}$, where $\alpha_1, \alpha_2 > 0$. In practice, we take $\alpha_1 = \alpha_2 = 1$, i.e., the Uniform distribution, as our default prior on $\boldsymbol{\omega}$. Besides this selection, we could also choose other $\alpha_1, \alpha_2$ values so as to emphasize or discount the historical control information in the prior. For example, if we set $\alpha_1$ to a large value and make $\alpha_2$ small, then significant weight is placed on the historical control information \emph{a priori}. In addition, if we set $\alpha_1$ to be small and make $\alpha_2$ large, then the prior will significantly discount the historical control information.

\subsubsection{Hyperparameter Specifications} \label{sec:hyperparameters}

Hyperparameter specification for the informative component is a key step in the method and has direct consequences on its operating characteristics. Some hyperparameters can be set directly from historical data, such that the informative prior component is directly interpreted and justified according to the posterior from the historical control data. Others may be tuned to discount historical information when, for example, there are discrepancies between historical and RCT data, so as to limit bias and variance inflation in inferences.

The hyperparameter values $\widehat{\beta_{0,H}}, \widehat{\beta_{2,H}}, s_H^2$, and the shape parameter $N_H-2$ are directly selected based on the posterior distribution of $\left ( \beta_0, \beta_2, \sigma^2 \right )^{\mathsf{T}}$ when the Bayesian linear regression model is fit for the $y_{i,H}^{(c)}$ on the $m_{i,H} - \bar{m}_H$ using the standard non-informative prior. The values for $K_{0,H}, K_{1,H}, K_{2,H}$ can be specified based on historical control data, and optionally tuned to discount historical information. For example, one option is to set $\begin{pmatrix} K_{0,H} & 0 \\ 0 & K_{2,H} \end{pmatrix} = \left ( V_H^{\mathsf{T}} V_H \right )^{-1}$ and $K_{1,H}$ to be a large value with respect to the scale of the endpoint. Keeping $K_{1,H}$ as a finite, large value ensures that $p_I \left ( \beta, \sigma^2 \right )$ is a generative and proper prior. By these definitions, $K_{0,H} = N_H^{-1}$, but this can lead to a posterior distribution that is excessively confident in the historical control information, which can yield biased inferences in cases of domain shifts between the historical and RCT data.

In cases of potential domain shifts between historical and trial populations, one can change the specification of $K_{0,H}, K_{2,H}$ such that historical information is discounted. Specifically, for a given amount of expected domain shift, one can set values for $K_{0,H}, K_{2,H}$ to control expected operating characteristics up to that amount of shift. In practice, an expected amount of domain shift can be selected using repeated sampling on a historical control dataset. This is further described in the Appendix \ref{appendix:v2}.

\subsection{Considerations for the Additive Mixture Prior}

The additive mixture prior in Bayesian PROCOVA has several advantages. First, it is easy to explain and justify. The distribution for $\left ( \beta_0, \beta_2, \sigma^2 \right )^{\mathsf{T}}$ in the informative prior component is the posterior distribution of the parameters from the historical control data. Hence, Bayesian PROCOVA is, in part, updating the historical control posterior with data from the RCT to calculate a new posterior for all parameters. In addition, as both the informative and flat prior components are proper probability distributions, the additive mixture prior is guaranteed to be a proper probability distribution. Second, the additive mixture prior for Bayesian PROCOVA is structured such that it is straightforward to understand the encoding of historical control information, and how that information can be discounted to make the informative prior component less dominant. Besides the point estimates $\widehat{\beta_{0,H}}, \widehat{\beta_{2,H}}$, and $s_H^2$, the values of $K_{0,H}$ and the prior shape parameter $N_H-2$ in the informative component are functions of the historical control sample size, and these two quantities can be decreased to discount the historical control information. The values of $\alpha_1$ and $\alpha_2$ can also be set to further discount the historical control information in the prior. This tuning is helpful in cases of domain shift to control bias while maintaining precision gains over PROCOVA. Third, by taking $\boldsymbol{\omega} \rightarrow 0$, $k \rightarrow \infty$, and $\nu_0 \rightarrow 0$, the additive mixture prior will converge to the standard non-informative prior $p \left ( \beta, \sigma^2 \right ) \propto \left ( \sigma^2 \right )^{-1}$, and hence Bayesian PROCOVA will yield treatment effect inferences similar to those from PROCOVA.

The amount of information contained in the prior of $\beta_0$ and $\beta_1$ conditional on a value of $\boldsymbol{\omega}$, i.e., $p \left ( \beta_0, \beta_1 \mid \boldsymbol{\omega} \right )$, under Bayesian PROCOVA can be quantified by comparing the prior variances of $\beta_0$ and $\beta_1$ from Bayesian PROCOVA to the posterior variances of those parameters that would be obtained if the RCT data were analyzed using a Bayesian linear regression model with predictor vector $v_i = \left ( 1, w_i \right )^{\mathsf{T}}$ and prior $p \left ( \beta_0, \beta_1, \sigma^2 \right ) \propto \left ( \sigma^2 \right )^{-1}$. By setting the prior variances of $\beta_0$ and $\beta_1$ conditional on $\boldsymbol{\omega}$ under Bayesian PROCOVA equal to their posterior variances in the latter, hypothetical Bayesian analysis, we can leverage closed-form expressions for the posterior variances of the parameters to identify the sample size for the hypothetical RCT such that the amount of information provided by the RCT on the parameters would be equivalent to the amount of information on the parameters encoded in the prior from Bayesian PROCOVA. We condition on $\boldsymbol{\omega}$ for interpreting the prior effective sample size in this regard as it can be used for both trial planning purposes and sensitivity analyses.

To illustrate this approach, first consider $\beta_1$. The calculation of $\mathrm{Var} \left ( \beta_1 \mid \boldsymbol{\omega} \right )$ under Bayesian PROCOVA follows in a straightforward manner due to the additive mixture prior, or alternatively via simulation as the prior is generative. For a hypothetical RCT with $N_1$ treated participants and $N - N_1 = N_0$ control participants in which the centered outcomes $y_i^{(c)}$ are analyzed using Bayesian linear regression with $v_i = \left ( 1, w_i \right )^{\mathsf{T}}$ and in which $p \left ( \beta_0, \beta_1, \sigma^2 \right ) \propto \left ( \sigma^2 \right )^{-1}$, the posterior variance of $\beta_1$ is $s^2 \left ( N_1^{-1} + N_0^{-1} \right )$ where $s^2$ is the estimate of $\sigma^2$ from the regression model. If we were to consider 1:1 designs, with $N_0 = N_1 = N/2$, then we set $\mathrm{Var} \left ( \beta_1 \mid \boldsymbol{\omega} \right ) = 4s^2N^{-1}$ and solve for $N$ to obtain $N = 4s^2/\mathrm{Var} \left ( \beta_1 \mid \boldsymbol{\omega} \right )$. Thus, given the estimate $s^2$ of $\sigma^2$ and a value of $\boldsymbol{\omega}$, the amount of prior information on $\beta_1$ conditional on $\boldsymbol{\omega}$ from Bayesian PROCOVA is equivalent to the corresponding amount of information in the posterior distribution for $\beta_1$ that is obtained from analyzing the hypothetical RCT of size $N = 4s^2/\mathrm{Var} \left ( \beta_1 \mid \boldsymbol{\omega} \right )$ using the simpler Bayesian linear regression analysis. The existence of $\mathrm{Var} \left ( \beta_1 \mid \boldsymbol{\omega} \right )$ for this calculation is ensured because Bayesian PROCOVA uses proper, generative prior components that have finite first and second moments. The same approach can be implemented for quantifying the information on $\beta_0$ from Bayesian PROCOVA.

\subsection{Posterior Distributions}
\label{sec:posterior_distributions}

The calculation of $p \left ( \beta, \sigma^2, \boldsymbol{\omega} \mid y, w, X \right )$ in Bayesian PROCOVA is performed in a straightforward manner. In particular, the combination of the additive mixture prior with the likelihood results in $p \left ( \beta, \sigma^2 \mid \boldsymbol{\omega}, y, w, X \right )$ being a mixture distribution itself with two components, corresponding to the informative and flat components in the prior. For each mixture component, the specification of a Multivariate Normal distribution for the conditional prior $\left [ \beta \mid \sigma^2 \right ]$ and the Inverse Chi-Square distribution for the marginal prior of $\sigma^2$ is conjugate to the likelihood function. As such, for both the informative and flat components, the conditional posterior $\left [ \beta \mid \sigma^2, y, w, X \right ]$ is a Multivariate Normal distribution, and the marginal posterior $\left [ \sigma^2 \mid y, w, X \right ]$ is an Inverse Chi-Square distribution. Both the marginal posterior $p \left ( \boldsymbol{\omega} \mid y, w, X \right )$ and the conditional posterior $p \left ( \boldsymbol{\omega} \mid \beta, \sigma^2, y, w, X \right )$ for the mixture weight can be derived in closed-form, with the normalizing constant calculated via numerical integration over the support $(0,1)$. The latter conditional posterior is typically more numerically stable than the former marginal posterior. These observations indicate a straightforward Gibbs algorithm for sampling from the joint posterior, with the algorithm alternating between sampling from $p \left ( \beta, \sigma^2 \mid \boldsymbol{\omega}, y, w, X \right )$ and from $p \left ( \boldsymbol{\omega} \mid \beta, \sigma^2, y, w, X \right )$. 

The fact that $p \left ( \beta, \sigma^2 \mid \boldsymbol{\omega}, y, w, X \right )$ is a mixture distribution is evident from Bayes' theorem, as
\begin{equation*}
p \left ( \beta, \sigma^2\mid \boldsymbol{\omega}, y, w, X \right ) \propto \boldsymbol{\omega} L \left ( \beta, \sigma^2, \boldsymbol{\omega} \mid y, w, X \right ) p_I \left ( \beta, \sigma^2\right ) + \left ( 1 - \boldsymbol{\omega} \right ) L \left ( \beta, \sigma^2, \boldsymbol{\omega} \mid y, w, X \right )p_F \left ( \beta, \sigma^2 \right ).
\end{equation*}
The normalization constant for this mixture distribution is
\begin{equation*}
C = \left \{ \boldsymbol{\omega} \displaystyle \int L \left ( \beta, \sigma^2, \boldsymbol{\omega} \mid y, w, X \right ) p_I \left ( \beta, \sigma^2 \right ) d\beta d\sigma^2 + \left ( 1 - \boldsymbol{\omega} \right ) \int L \left ( \beta, \sigma^2, \boldsymbol{\omega} \mid y, w, X \right ) p_F \left (\beta, \sigma^2 \right ) d \beta d \sigma^2 \right \}^{-1}.
\end{equation*}
The weights for the two components of the posterior are $\boldsymbol{\omega}_* = C \boldsymbol{\omega} \displaystyle \int L \left ( \beta, \sigma^2, \boldsymbol{\omega} \mid y, w, X \right ) p_I \left ( \beta, \sigma^2 \right ) d\beta d\sigma^2$ and $1 - \boldsymbol{\omega}_* = C \left ( 1 - \boldsymbol{\omega} \right ) \displaystyle \int L \left ( \beta, \sigma^2, \boldsymbol{\omega} \mid y, w, X \right ) p_F \left (\beta, \sigma^2 \right ) d \beta d\sigma^2$, respectively. Closed-form expressions for $C^{-1}, \boldsymbol{\omega}_*$, and $1 - \boldsymbol{\omega}_*$ are given in the Appendix.
For the informative component of the mixture conditional posterior, $\left [ \beta \mid \sigma^2, y, w, X \right ]$ is Multivariate Normal with mean vector $\beta_* = \left ( \beta_{0,*}, \beta_{1,*}, \beta_{2,*} \right )^{\mathsf{T}}$ and covariance matrix $\boldsymbol{\Sigma}_*$ defined as
\begin{align*}
\beta_* &= \begin{pmatrix} \widehat{\beta_{0,H}} \\ 0 \\ \widehat{\beta_{2,H}} \end{pmatrix} + \boldsymbol{K} V^{\mathsf{T}} \left ( I_{N \times N} + V \boldsymbol{K} V^{\mathsf{T}} \right )^{-1} \left ( y^{(c)} - V  \begin{pmatrix} \widehat{\beta_{0,H}} \\ 0 \\ \widehat{\beta_{2,H}} \end{pmatrix} \right ), \\
\boldsymbol{\Sigma}_* &= \sigma^2 \boldsymbol{K} - \sigma^2 \boldsymbol{K}V^{\mathsf{T}} \left ( I_{N \times N} + V \boldsymbol{K} V^{\mathsf{T}} \right )^{-1} V \boldsymbol{K}.
\end{align*}
The posterior of $\sigma^2$ in the informative component is $p_I \left ( \sigma^2 \mid y, w, X \right ) = p_I \left ( \beta, \sigma^2 \mid y, w, X \right )/p_I \left ( \beta \mid \sigma^2, y, w, X \right )$, so that $\left [ \sigma^2 \mid y, w, X \right ] \sim (N+N_H-2)s_*^2/\chi_{N+N_H-2}^2$ with
\begin{align*}
s_*^2 &= \left ( N+N_H-2 \right )^{-1} \Bigg{\{} \left ( y^{(c)} - V\beta_* \right )^{\mathsf{T}} \left (  y^{(c)} - V\beta_* \right ) + (N_H-2)s_H^2 \\
& \ \ \ \ \ \ \ \ \ \ \ \ \ \ \ \ \ \ \ \ \ \ \ \ \ \ \ \ + \frac{ \left ( \beta_{0,*} - \widehat{\beta_{0,H}} \right )^2}{K_{0,H}} + \frac{\beta_{1,*}^2}{K_{1,H}} + \frac{ \left ( \beta_{2,*} - \widehat{\beta_{2,H}} \right )^2}{K_{2,H}} \Bigg{\}}.
\end{align*}

For the flat component of the mixture conditional posterior, $\left [ \beta \mid \sigma^2, y, w, X \right ]$ is Multivariate Normal with mean vector $b_*= \left ( b_{0,*},  b_{1,*},  b_{2,*} \right )^{\mathsf{T}}$ and covariance matrix $\mathbf{S}_*$ defined as
\begin{align*}
    b_* &= kV^{\mathsf{T}} \left ( I_{N \times N} + kVV^{\mathsf{T}} \right )^{-1} y^{(c)}, \\
    \mathbf{S}_* &= \sigma^2 kI_{3 \times 3} -  \sigma^2 k^2 V^{\mathsf{T}} \left ( I_{N \times N} + kVV^{\mathsf{T}} \right )^{-1} V.
\end{align*}
Similar to the previous case, the marginal posterior of $\sigma^2$ in the flat component is $\left [ \sigma^2 \mid y, w, X \right ]\sim (N+\nu_0)\sigma_{0,*}^2/\chi_{N+\nu_0}^2$, where
\[
\sigma_{0,*}^2 = \left ( N+\nu_0 \right )^{-1} \left \{ \left ( y^{(c)} - Vb_* \right )^{\mathsf{T}} \left (  y^{(c)} - Vb_* \right ) + \nu_0\sigma_0^2 + \frac{1}{k} \left( b_{0,*}^2 + b_{1,*}^2 + b_{2,*}^2 \right) \right \}.
\]
Formulae for the posterior mean and variance of $\beta_1$ conditional on $\boldsymbol{\omega}$ are derived via our previous expressions for the posteriors of the parameters. The posterior mean of $\beta_1$ is the second entry in
\begin{align*}
\mathbb{E} \left ( \beta \mid \boldsymbol{\omega}, y, w, X \right ) &= \boldsymbol{\omega}_* \beta_* + (1 - \boldsymbol{\omega}_*) b_*.
\end{align*}
The posterior variance of $\beta_1$ is calculated according to the law of total variance via
\begin{align*}
\mathrm{Cov} \left ( \beta \mid \boldsymbol{\omega}, y, w, X \right ) &= \boldsymbol{\omega}_* \left \{ \frac{\left ( N + N_H - 2 \right )s_*^2}{N+N_H-4} \right \} \Bigg{\{} \boldsymbol{K} - \boldsymbol{K} V^{\mathsf{T}} \left ( I_{N \times N} + V \boldsymbol{K} V^{\mathsf{T}} \right )^{-1} V \boldsymbol{K} \Bigg{\}} \\
& \ \ \ + \left ( 1 - \boldsymbol{\omega}_* \right ) \left \{ \frac{\left ( N + \nu_0 \right) \sigma_{0,*}^2}{N+\nu_0-2} \right \} \left \{ kI_{3 \times 3} - k^2 V^{\mathsf{T}} \left ( I_{N \times N} + kVV^{\mathsf{T}} \right )^{-1} V \right \} \\
& \ \ \ + \boldsymbol{\omega}_* \left ( 1 - \boldsymbol{\omega}_* \right )\left( \beta_* - b_* \right) \left( \beta_* - b_* \right)^{\mathsf{T}}.
\end{align*}
The posterior variance of $\beta_1$ is the $(2,2)$ entry of this matrix. By means of additional algebra, this formula indicates that the variance reduction of Bayesian PROCOVA over PROCOVA depends in part on comparisons between the historical control and RCT data in regard to the correlation between the prognostic scores and outcomes and the average bias in the prognostic scores in both datasets. The variance reduction is also a function of the sample sizes of the historical control and RCT data. These insights can help guide sensitivity analyses for Bayesian PROCOVA. 

The \citet[p.~40]{food_and_drug_administration_bayesian_2010} provided a simple formula for the effective sample size (ESS) of a Bayesian analysis. This formula is $\mathrm{ESS} = N(\mathbb{V}_1/\mathbb{V}_2)$, where $\mathbb{V}_1$ is the posterior variance of the parameter of interest (e.g., $\beta_1$) that is obtained from an analysis without using an informative prior (i.e., PROCOVA under the standard non-informative prior), and $\mathbb{V}_2$ is the posterior variance of the same parameter under a more informative prior (i.e., Bayesian PROCOVA). We have closed-form expressions for variances conditional on $\boldsymbol{\omega}$, and can calculate the ESS accordingly. Furthermore, we can calculate the sample size reduction via $\mathrm{ESS} - N = N(\mathbb{V}_1/\mathbb{V}_2 - 1)$. This formula is also present in the work of \citet[p.~177]{kaizer_2018}, \citet{hobbs_2013}, and \citet{han_2017}. A more complicated expression for ESS was given by \citet{zhao_2023}, but the essential ingredient in their expression was the formula above.  

We calculate the marginal posterior $p \left ( \boldsymbol{\omega} \mid y, w, X \right ) = p \left ( \beta, \sigma^2, \boldsymbol{\omega} \mid y, w, X \right )/p \left ( \beta, \sigma^2 \mid \boldsymbol{\omega}, y, w, X \right )$. As this is a function of $\boldsymbol{\omega}$, we can input any set of values for $\beta$ and $\sigma^2$ into the right-hand side of the equation to obtain $p \left ( \boldsymbol{\omega} \mid y, w, X \right )$. We take care to include all normalizing constants in the numerator and denominator that involve $\boldsymbol{\omega}$ in this calculation. This is a one-dimensional distribution whose support is on $(0,1)$, and so we can calculate the normalizing constant of this marginal posterior using numerical integration. In addition, the posterior of $\boldsymbol{\omega}$ conditional on $\beta, \sigma^2$ is directly calculated as
\begin{align*}
p \left ( \boldsymbol{\omega} \mid \beta, \sigma^2, y, w, X \right ) &\propto \boldsymbol{\omega} p \left ( \boldsymbol{\omega} \right ) \\
& \ \ \ \times \left \{ \frac{ \left ( \frac{N_H - 2}{2}\right )^{\left ( N_H-2 \right )/2} \left ( s_H^2 \right )^{\left ( N_H-2 \right )/2} \left ( \sigma^2 \right )^{-\left \{ (N_H+1)/2 + 1 \right \}}}{\Gamma \left ( \frac{N_H-2}{2} \right ) \pi^{3/2}  \left (K_{0,H} K_{1,H} K_{2,H} \right )^{1/2}} \right\} \\
& \ \ \ \times \mathrm{exp} \left [ -\frac{ \left ( N_H - 2 \right ) s_H^2}{2\sigma^2} - \frac{1}{2\sigma^2} \left \{ \frac{\left ( \beta_0 - \widehat{\beta_{0,H}} \right )^2}{K_{0,H}} + \frac{\beta_1^2}{K_{1,H}} + \frac{ \left ( \beta_2 - \widehat{\beta_{2,H}} \right )^2}{K_{2,H}} \right \} \right ] \\
& \ \ \ + \left ( 1 - \boldsymbol{\omega} \right ) p \left ( \boldsymbol{\omega} \right ) \\
& \ \ \ \times \left \{ \frac{ \left ( \frac{\nu_0}{2} \right )^{\nu_0/2} \left ( \sigma_0^2 \right )^{\nu_0/2} \left ( \sigma^2 \right )^{-\left \{ (\nu_0 + 3)/2 +1 \right \}}}{\Gamma \left ( \frac{\nu_0}{2} \right ) \left ( \pi k \right)^{3/2}} \right \} \mathrm{exp} \left \{ -\frac{\nu_0 \sigma_0^2}{2\sigma^2} - \frac{1}{2k\sigma^2} \left ( \beta_0^2 + \beta_1^2 + \beta_2^2 \right ) \right \}.
\end{align*}
Similar to the marginal posterior, for any $\beta$ and $\sigma^2$ we can calculate the normalizing constant of this conditional distribution using numerical integration, and thereby directly obtain samples from it.

\subsection{Gibbs Sampler for the Mixture Posterior Distribution}
\label{sec:gibbs_sampler}

We sample from the mixture posterior $p \left ( \beta, \sigma^2, \boldsymbol{\omega} \mid y, w, X \right )$ using a Gibbs algorithm. Specifically, we iterate between drawing the vector of parameters from $p \left ( \beta, \sigma^2 \mid \boldsymbol{\omega}, y, w, X \right )$ conditional on a previous draw of $\boldsymbol{\omega}$, and drawing $\boldsymbol{\omega}$ from $p \left ( \boldsymbol{\omega} \mid \beta, \sigma^2, y, w, X \right )$ conditional on the previously drawn $\beta$ and $\sigma^2$. The formal steps for the Gibbs algorithm are outlined below.

\begin{enumerate}

\item[0.] Initialize $\boldsymbol{\omega}^{(0)}$.

For iteration $j = 1, 2, \ldots$:

\item Calculate $\boldsymbol{\omega}_*$ based on $\boldsymbol{\omega}^{(j-1)}$.

\item Draw $Z^{(j)} \sim \mathrm{Bernouilli} \left ( \boldsymbol{\omega}_* \right )$.

\item If $Z^{(j)} = 1$:

\begin{enumerate}

\item Draw $\left ( \sigma^2 \right )^{(j)} \sim \left \{ (N+N_H-2)s_*^2 \right \} / \chi_{N+N_H-2}^2$.

\item Draw $\beta^{(j)}$ from the informative posterior component $\left [ \beta \mid \left ( \sigma^2 \right )^{(j)}, y, w, X \right ]$.

\end{enumerate}

\item If $Z^{(j)} = 0$:

\begin{enumerate}
    
\item Draw $\left ( \sigma^2 \right )^{(j)} \sim \left \{ (N+\nu_0)\sigma_{0,*}^2 \right \} / \chi_{N+\nu_0}^2$.

\item Draw  $\beta^{(j)}$ from the flat posterior component $\left [ \beta \mid \left ( \sigma^2 \right )^{(j)}, y, w, X \right ]$.

\end{enumerate}

\item Draw $\boldsymbol{\omega}^{(j)}$ via the Probability Integral Transform applied to the cumulative distribution function of $p \left ( \boldsymbol{\omega} \mid \beta^{(j)}, \left ( \sigma^2 \right )^{(j)}, y, w, X \right )$ as obtained by numerical integration.

\end{enumerate}

\section{Simulation Studies}
\label{sec:simulation_experiments}

\subsection{Data Generation Mechanisms and Evaluation Metrics}
\label{sec:dgm_metrics}

We conduct simulation studies under five scenarios to evaluate the frequentist performance, in terms of bias control and variance reduction, of Bayesian PROCOVA compared to PROCOVA. In the first three scenarios, the historical control and RCT data are generated according to the same mechanism. Here we explore the sensitivity of the frequentist properties of Bayesian PROCOVA to the prior on $\boldsymbol{\omega}$ (which will either be the Beta$(1,1)$ or the Beta$(1/2,1/2)$ distribution) and the prior on $\sigma^2$ in the flat prior component (which will have either $\nu_0 = 1, \sigma_0^2 = 1$ or $\nu_0 = 3, \sigma_0^2 = 100$). In the fourth scenario, we introduce shifts in the correlation of the prognostic scores with the control outcomes between the historical control and RCT data. In this case we only consider the Beta$(1,1)$ prior for $\boldsymbol{\omega}$ and the Inverse Chi-Square prior with $\nu_0 = 1, \sigma_0^2 = 1$ for $\sigma^2$ in the flat prior component. Lastly we introduce shifts in the average bias of the prognostic score in the historical control data compared to the RCT. Here we only consider the Beta$(1,1)$ prior for $\boldsymbol{\omega}$, and the Inverse Chi-Square priors for $\sigma^2$ with $\nu_0 = 1, \sigma_0^2 = 1$ and $\nu_0 = 3, \sigma_0^2 = 100$ in the flat prior component (Table \ref{tab:varied-sim-params}). 

Table \ref{tab:fixed-sim-params} gives the parameter settings that remain fixed across simulation scenarios. Observed outcomes in both historical and trial datasets are generated according to equation (\ref{eq:bayesian_procova_model}) with prognostic scores generated identically and independently from one another based on standard Normal random variables. In the historical data, prognostic scores are on average unbiased ($\beta_{0,H} = 0$) and error variance is set as $\sigma^2  =1$. In the trial data, participants are randomized at a $1:1$ ratio for the trial and a null treatment effect ($\beta_1=0$). The value of $\beta_0$ in the trial data depends on the shift in bias from the historical data, which we vary in one simulation scenario. The value of $\beta_2$ in both the historical and trial datasets depend on the correlation between the prognostic scores and control outcomes. Multiple levels of correlation are considered in each scenario, and in one scenario we introduce discrepancies between the values in the historical and trial data. All scenarios are simulated for both the case of $K_{0,H} = N_H^{-1}$ and $K_{0,H} = N_H^{-1/2}$. For each scenario, $1000$ datasets are simulated, and the Gibbs algorithm is implemented for $1000$ iterations in each simulated dataset to fit the Bayesian PROCOVA model. We confirmed that the Gibbs algorithm converged rapidly, and validated the implementation of the Gibbs algorithm using the diagnostics of \citet{cook_2006}. The metrics that we calculate across the simulated datasets are bias and variance reduction with respect to PROCOVA. 

\begin{table}[H]
    \centering
    \resizebox{\columnwidth}{!}{
    \begin{tabular}{|l|c|c|c|c|c|c|}
    \hline
       \textbf{Factor} & \textbf{Notation} & \textbf{Scenario 1} & \textbf{Scenario 2} & \textbf{Scenario 3} & \textbf{Scenario 4} & \textbf{Scenario 5} \\ 
       \hline \hline
       Prior on weight $\omega$ & Beta$(\alpha_1,\alpha_2)$ & Beta(1,1) & \textcolor{red}{Beta(1/2,1/2)} & Beta(1,1) & Beta(1,1) & Beta(1,1) \\
       \hline
       Prior on $\sigma^2$ & InvChiSq($\nu_0,\sigma_0^2)$ & InvChiSq(1,1) & InvChiSq(1,1) & \textcolor{red}{InvChiSq(3,100)} & InvChiSq(1,1) & InvChiSq(1,1) \\
       \hline
       Correlation shift & $\rho_{m,y,T} - \rho_{m,y,H}$ & 0 & 0 & 0 & \textcolor{red}{-0.2, -0.1, 0.1, 0.2} & 0  \\
       \hline
       Bias shift &
       $\beta_{0,T} - \beta_{0,H}$ & 0 & 0 & 0 & 0 & \textcolor{red}{1, 2, 3, 4, 5} \\
       \hline
    \end{tabular}
    }
    \caption{Varied simulation parameters. Red text indicates parameter(s) that differ from the baseline settings of Scenario 1.}
    \label{tab:varied-sim-params}
\end{table}

\begin{table}[H]
    \centering
    \resizebox{\columnwidth}{!}{
    \begin{tabular}{|l|c|c|}
    \hline
       \textbf{Factor} & \textbf{Notation} & \textbf{Value} \\ 
       \hline \hline
       Trial sample size & $N$ & 25, 50, 100, 250 \\
       \hline
       Historical sample size & $N_H$ & 100, 300, 500 \\
       \hline
       Intercept term (historical data) & $\beta_{0,H}$ & 0 \\
       \hline
       Treatment effect & $\beta_1$ & 0 \\
       \hline
       Outcome variance (both historical and trial data) & $\sigma^2$ & 1 \\
       \hline
       Correlation between prognostic scores and observed outcomes in the historical data & $\rho_{m,y,H}$ & 0, 0.1, 0.2, 0.3, 0.4, 0.5 \\
       (the correlation in the RCT data is determined by the shift in Table \ref{tab:varied-sim-params}) & & \\
       \hline
       Randomization probability to treatment & $\pi$ & 0.5 \\
       \hline
       Positive constant for prior variance on $\beta_0$ in the informative component & $K_{0,H}$ & $N_H^{-1}$ and $N_H^{-1/2}$ \\
       \hline
       Positive constants for other prior variances & $K_{1,H}, k$ & 100,100\\
       \hline
    \end{tabular}
    }
    \caption{Fixed parameters across all simulation scenarios.}
    \label{tab:fixed-sim-params}
\end{table}

\subsection{Bias Control}
\label{sec:bias_control}

\begin{figure}[H]
\centering
\includegraphics[scale=0.2]{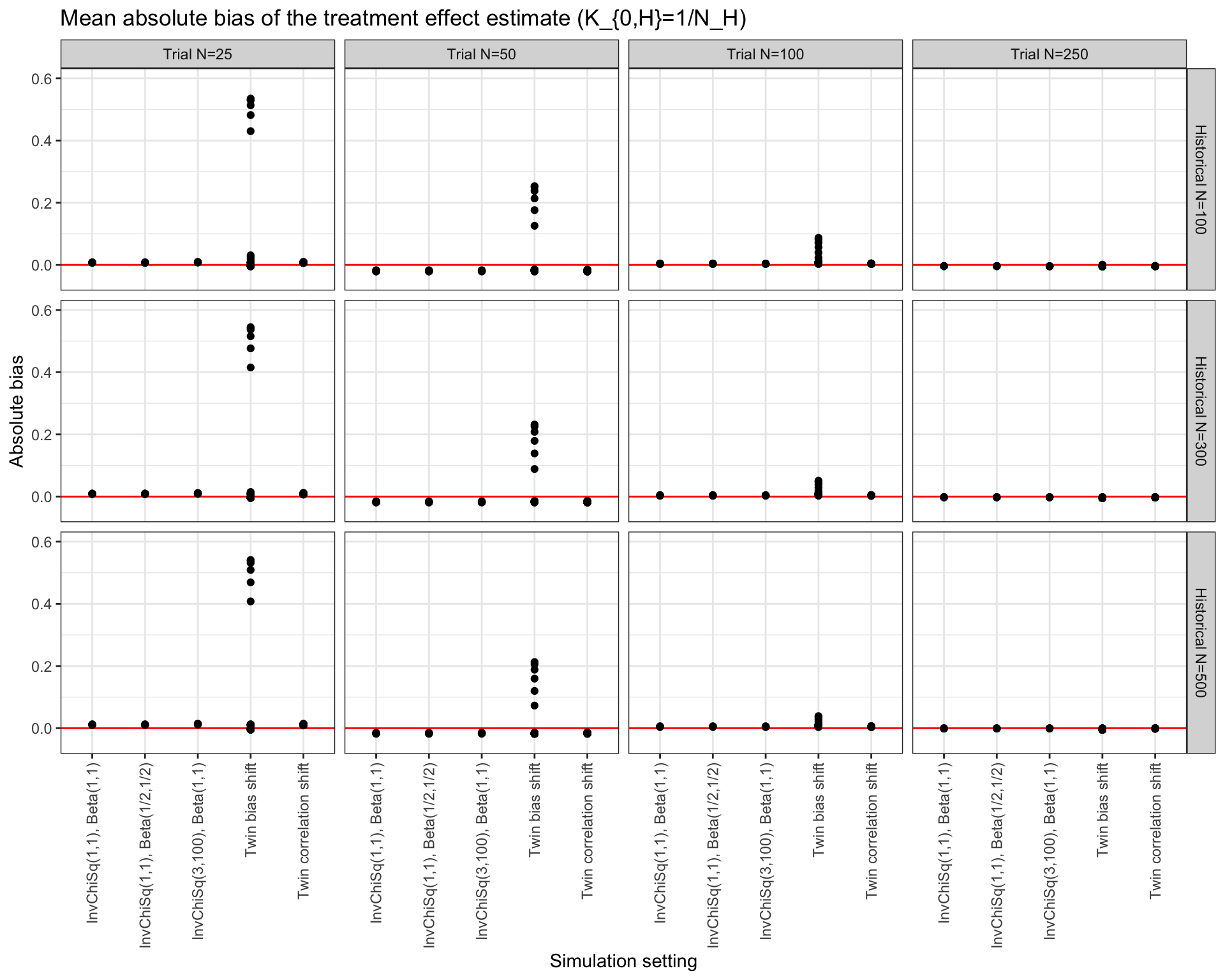}
\caption{Using $K_{0,H}=N_H^{-1}$: Mean absolute bias of the treatment effect estimator from Bayesian PROCOVA. The simulation settings are indicated on the horizontal axis, and the mean absolute bias of the treatment effects estimators are indicated on the vertical axis. A horizontal red line is included in the figures to represent zero bias. The individual points correspond to different parameter combinations within each historical-trial sample size pair (i.e., different correlation levels).}
\label{fig:bias_control}
\end{figure}

\begin{figure}[H]
\centering
\includegraphics[scale=0.2]{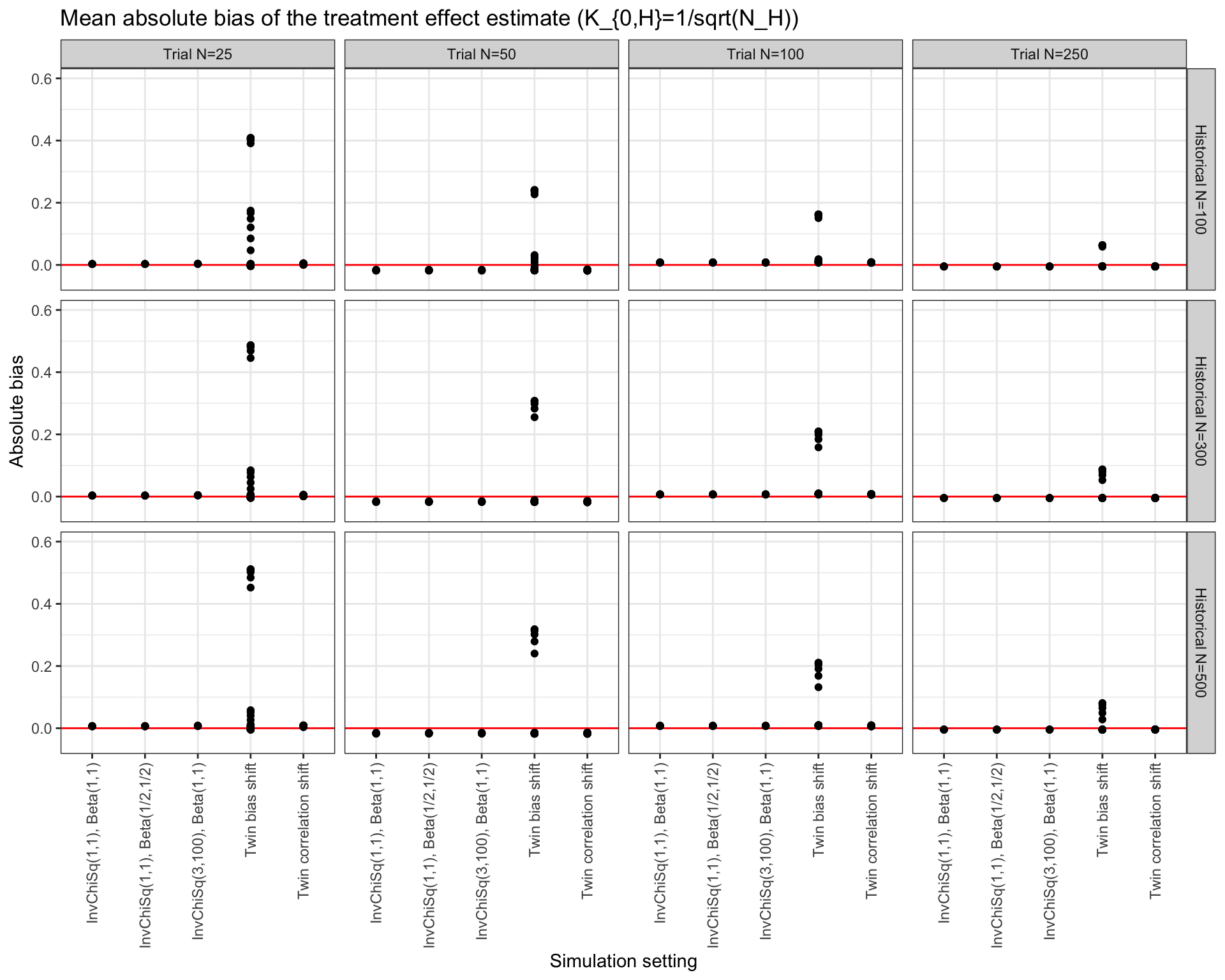}
\caption{Using $K_{0,H}=N_H^{-1/2}$: Mean absolute bias of the treatment effect estimator from Bayesian PROCOVA. The simulation settings are indicated on the horizontal axis, and the mean absolute bias of the treatment effects estimators are indicated on the vertical axis. A horizontal red line is included in the figures to represent zero bias. The individual points correspond to different parameter combinations within each historical-trial sample size pair (i.e., different correlation levels).}
\label{fig:bias_control-sqrt}
\end{figure}

\begin{figure}[H]
\centering
\includegraphics[scale=0.2]{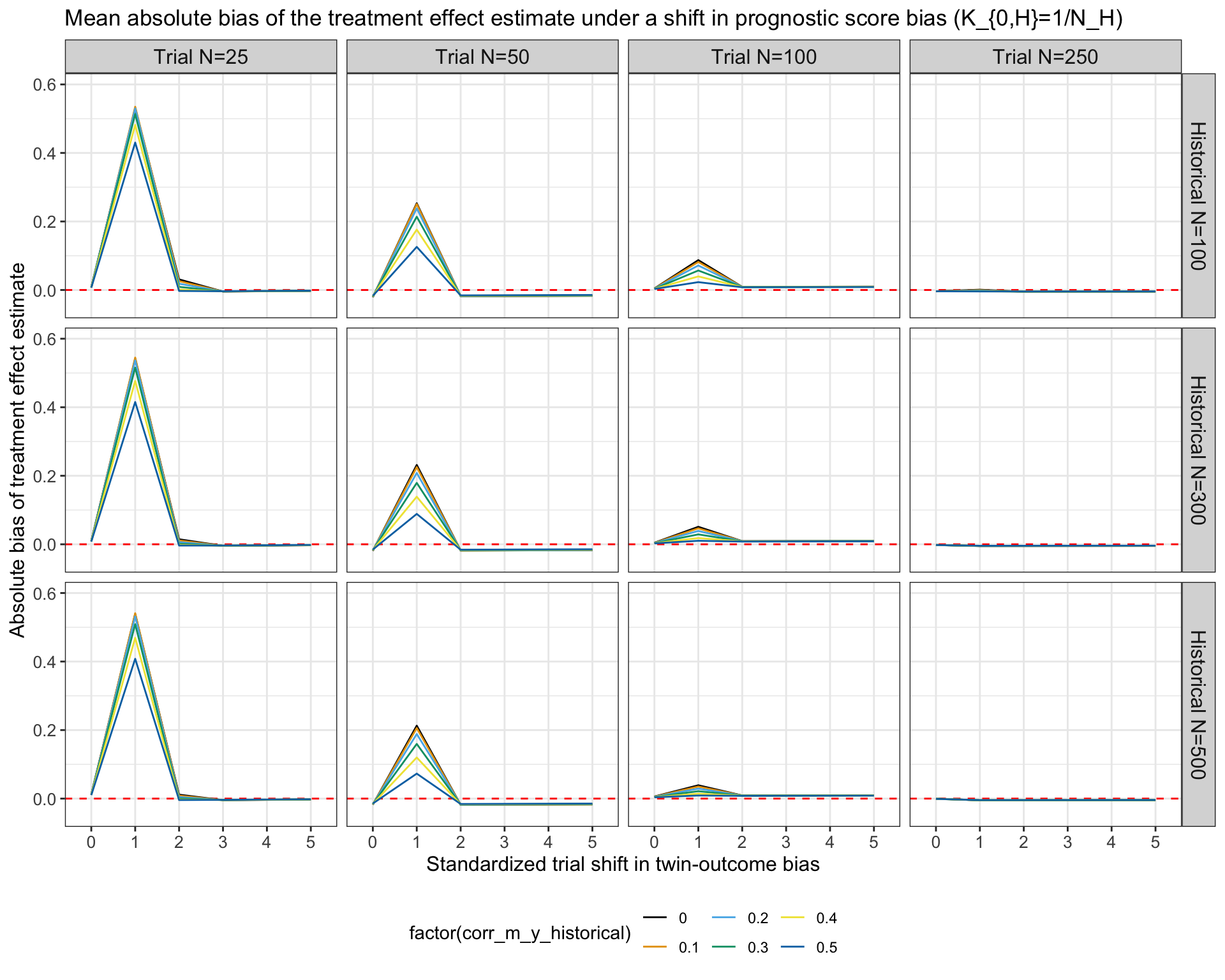}
\caption{Using $K_{0,H}=N_H^{-1}$: Mean absolute bias of the treatment effect estimator from Bayesian PROCOVA in the case of a shift in the mean bias ($\beta_0$) of the prognostic scores between historical and trial datasets. The standardized shift in $\beta_0$ is given on the horizontal axis, and the mean absolute bias of the treatment effects estimators are indicated on the vertical axis. A horizontal red line is included in the figures to represent zero bias. The individual points correspond to different correlation levels.}
\label{fig:bias_control_shift_beta0}
\end{figure}

\begin{figure}[H]
\centering
\includegraphics[scale=0.2]{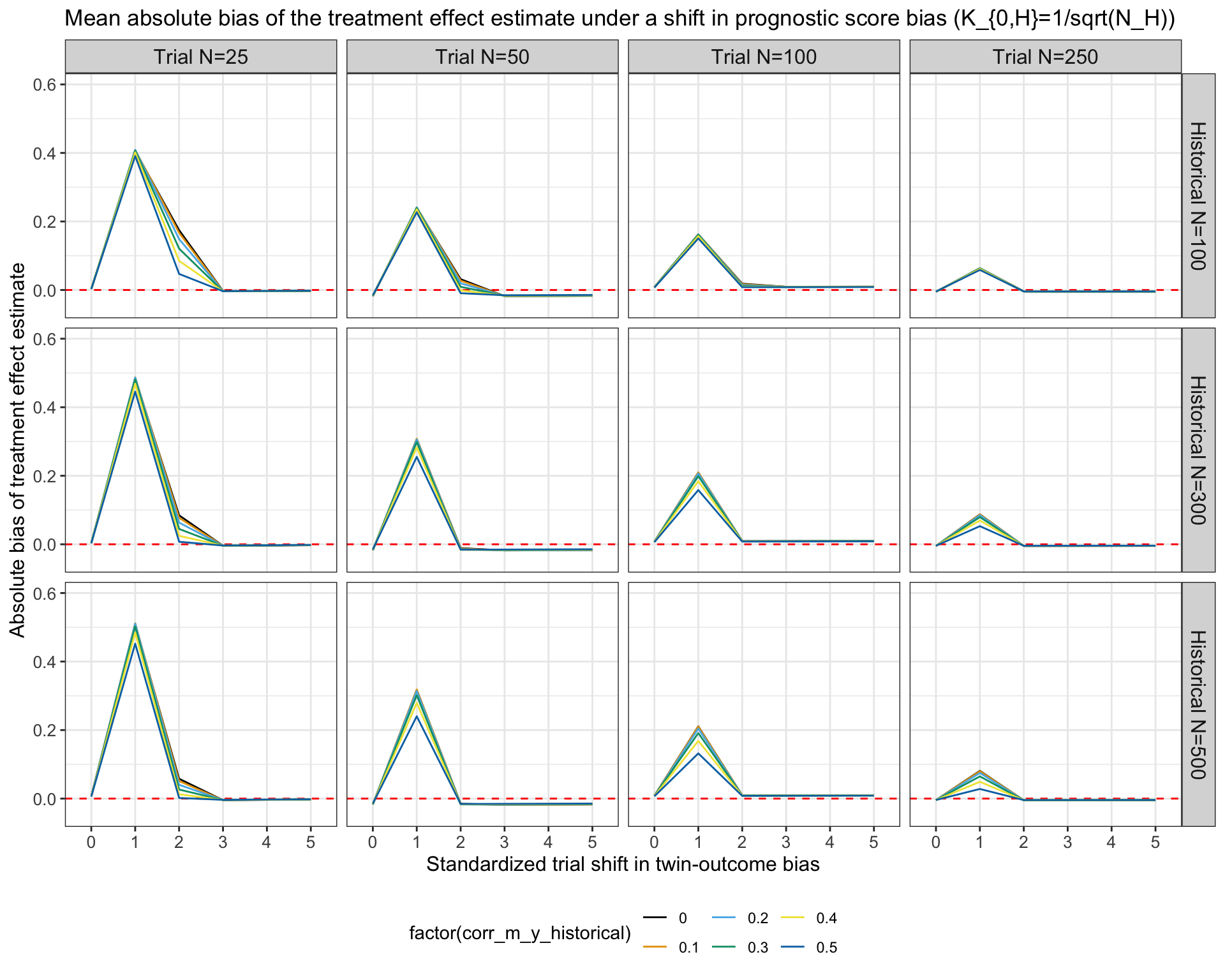}
\caption{Using $K_{0,H}=N_H^{-1/2}$: Mean absolute bias of the treatment effect estimator from Bayesian PROCOVA in the case of a shift in the mean bias ($\beta_0$) of the prognostic scores between historical and trial datasets. The standardized shift in $\beta_0$ is given on the horizontal axis, and the mean absolute bias of the treatment effects estimators are indicated on the vertical axis. A horizontal red line is included in the figures to represent zero bias. The individual points correspond to different correlation levels.}
\label{fig:bias_control_shift_beta0-sqrt}
\end{figure}

Figures \ref{fig:bias_control} to \ref{fig:bias_control_shift_beta0-sqrt} summarize the evaluations of the mean absolute bias of the posterior mean of the super-population treatment effect from Bayesian PROCOVA across different types of priors and simulation scenarios. We observe that bias in the Bayesian PROCOVA treatment effect estimator is effectively controlled in nearly all simulation scenarios (Figures \ref{fig:bias_control} and \ref{fig:bias_control-sqrt}). The exception is the situation in which the RCT has a small sample size and there exists a mild-to-moderate shift in the absolute prognostic score bias between the historical and RCT data (Figures \ref{fig:bias_control_shift_beta0} and \ref{fig:bias_control_shift_beta0-sqrt}). As the RCT sample size increases, the bias converges towards zero. By comparing the the top-left panels of 
Figures \ref{fig:bias_control_shift_beta0} and \ref{fig:bias_control_shift_beta0-sqrt}, we observe that setting $K_{0,H} = N_H^{-1/2}$ can control the maximum level of bias compared to setting $K_{0,H} = N_H^{-1}$ for a mild bias shift in the case of a small trial size. These two observations indicate that the bias is a consequence of the Bayesian method placing undue confidence in the historical control data as a result of the large historical control sample size and the relatively smaller RCT sample size. Additionally, in cases of prognostic score bias shift, the bias in the treatment effect estimator is smaller when the prognostic scores are more highly correlated with the control outcomes (Figures \ref{fig:bias_control_shift_beta0} and \ref{fig:bias_control_shift_beta0-sqrt}).

\subsection{Variance Reduction}
\label{sec:variance_reduction}

\begin{figure}[H]
\centering
\includegraphics[scale=0.2]{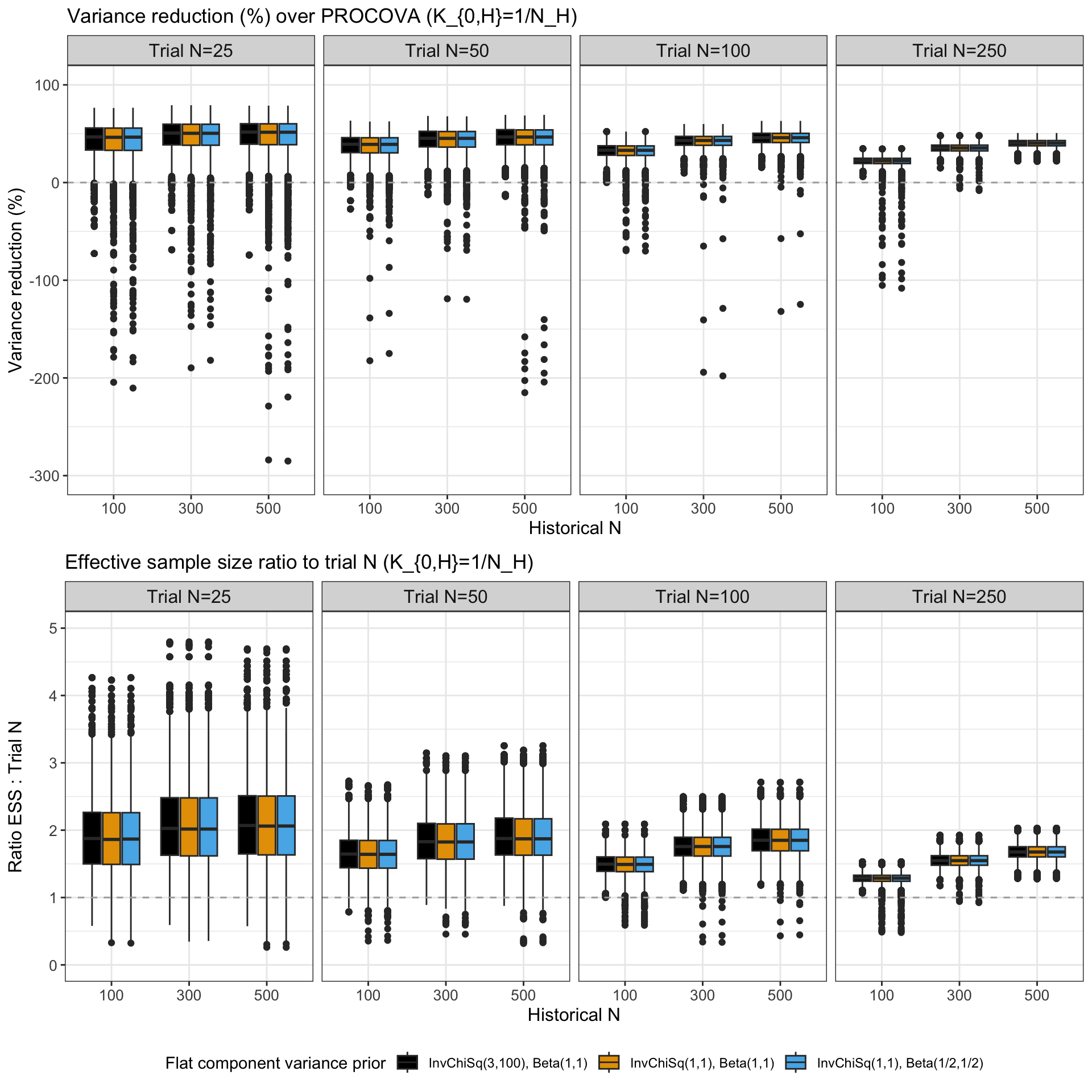}
\caption{Using $K_{0,H}=N_H^{-1}$: Boxplots across simulated datasets that demonstrate the variance reduction (in percentages) of Bayesian PROCOVA over PROCOVA (top row), and the ratio of the ESS to RCT sample size (bottom row). The colors indicate the hyperparameter settings. None of these settings involve discrepancies between the historical and RCT data. Whenever the ESS:$N$ ratio is less than one, Bayesian PROCOVA inflates variance compared to PROCOVA.}
\label{fig:variance_reduction}
\end{figure}

\begin{figure}[H]
\centering
\includegraphics[scale=0.2]{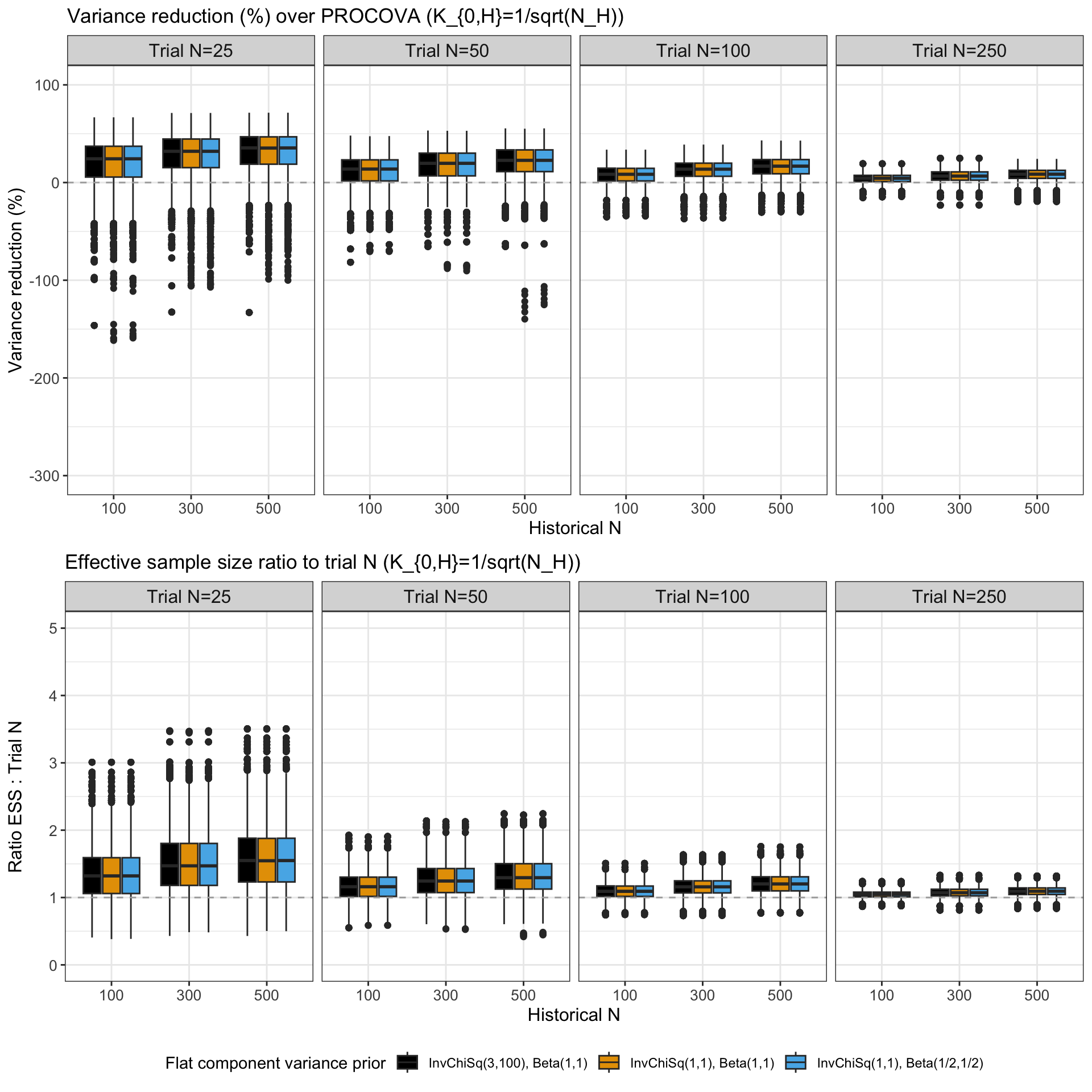}
\caption{Using $K_{0,H}=N_H^{-1/2}$: Boxplots across simulated datasets that demonstrate the variance reduction (in percentages) of Bayesian PROCOVA over PROCOVA (top row), and the ratio of the ESS to RCT sample size (bottom row). The colors indicate the hyperparameter settings. None of these settings involve discrepancies between the historical and RCT data. Whenever the ESS:$N$ ratio is less than one, Bayesian PROCOVA inflates variance compared to PROCOVA.}
\label{fig:varred_ess-sqrt}
\end{figure}

Figures \ref{fig:variance_reduction} and \ref{fig:varred_ess-sqrt} illustrate the variance reductions from Bayesian PROCOVA in the first situation for which the historical and RCT data are consistent with one another. We observe that the Inverse Chi-Square prior with $\nu_0 = 3, \sigma_0^2 = 100$ for $\sigma^2$ in the flat prior component and the Beta$(1,1)$ prior on $\boldsymbol{\omega}$ yield consistent and stable variance reduction for Bayesian PROCOVA over PROCOVA, even in small RCTs. Variance reduction is more unstable and smaller in expectation when the Inverse Chi-Square prior with $\nu_0 = 1, \sigma_0^2 = 1$ is utilized in the flat prior component. 

This result on variance reduction can be explained by considering how the flat prior component for $\sigma^2$ affects the posterior distribution of $\boldsymbol{\omega}$. Specifying the Inverse Chi-Square prior with $\nu_0 = 3, \sigma_0^2 = 100$ results in an average posterior weight of $\boldsymbol{\omega} \approx 1$, so that significant weight is placed on the information from the historical data. This is advantageous when the historical and RCT data are consistent with each other, because Bayesian PROCOVA effectively augments the small RCT sample size with all of the information in the larger historical control dataset. In contrast, the average posterior weight in the other case of $\nu_0 = 1, \sigma_0^2 = 1$ is $\boldsymbol{\omega} \approx 0.33$. This results in both less information being leveraged from the historical control data, and more of the weak information from the flat prior component being utilized in the posterior inferences. While bias in the treatment effect inferences remains under control in this circumstance, this mixture of information increases the variance and introduces additional instabilities in posterior inferences, especially for small RCT sample sizes. 

The posterior variance of the treatment effect is also directly related to the ESS. This relationship is displayed in Figures \ref{fig:variance_reduction} and \ref{fig:varred_ess-sqrt}. We see from these figures that, the larger the ratio of the ESS to the actual RCT sample size, the greater the variance reduction of Bayesian PROCOVA compared to PROCOVA. We also observe that this ratio decreases for larger RCTs.

\begin{figure}[H]
\centering
\includegraphics[scale=0.2]{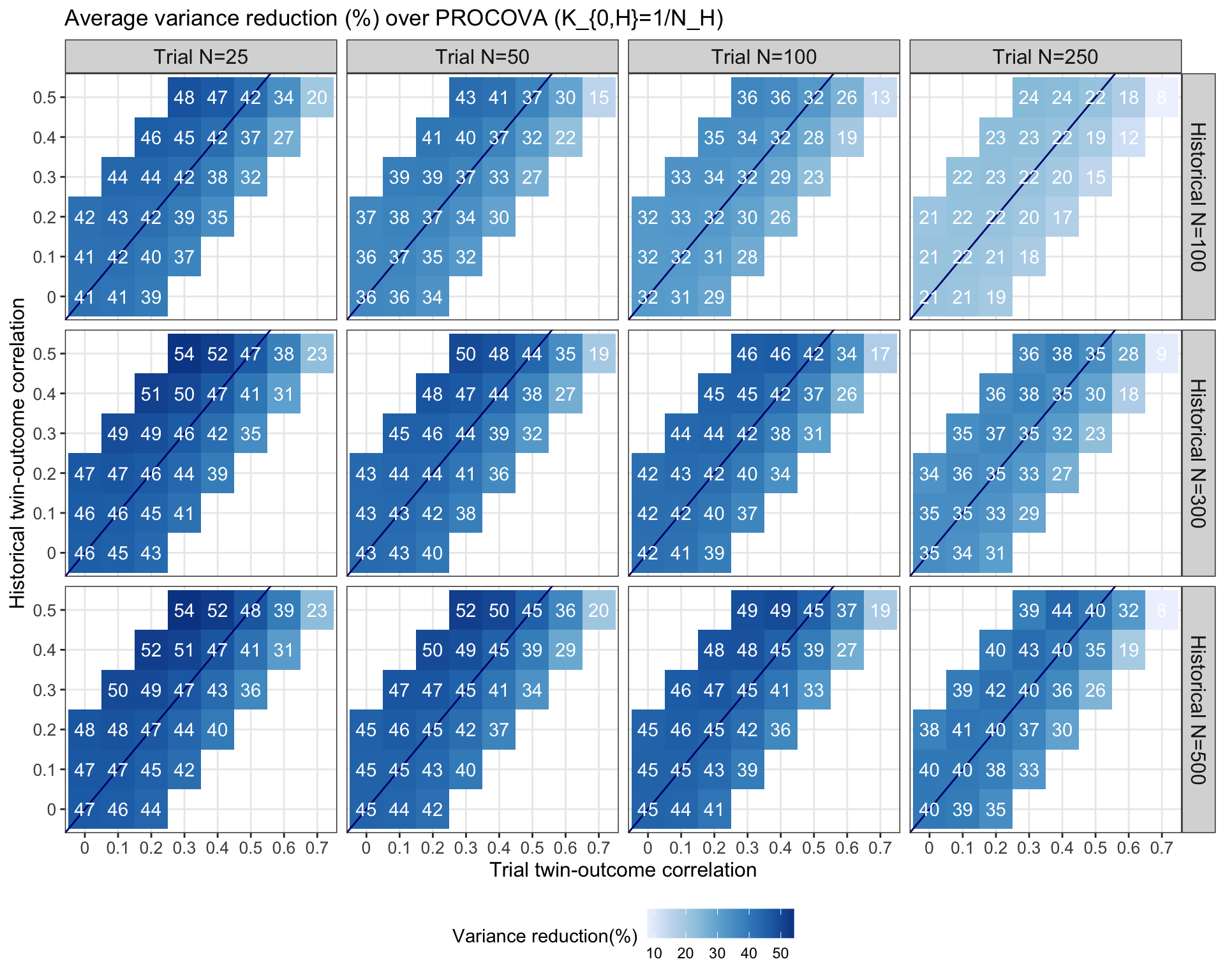}
\caption{Using $K_{0,H}=N_H^{-1}$: Average variance reduction (in percentages) of Bayesian PROCOVA over PROCOVA for different correlations of the prognostic scores with the control outcomes between historical and RCT data. The horizontal axis denotes the correlation in the RCT data and the vertical axis denotes the correlation in the historical data. The diagonal line in each panel represents cases in which the correlation is the same in both datasets.}
\label{fig:variance_reduction_correlation_shift}
\end{figure}

\begin{figure}[H]
\centering
\includegraphics[scale=0.2]{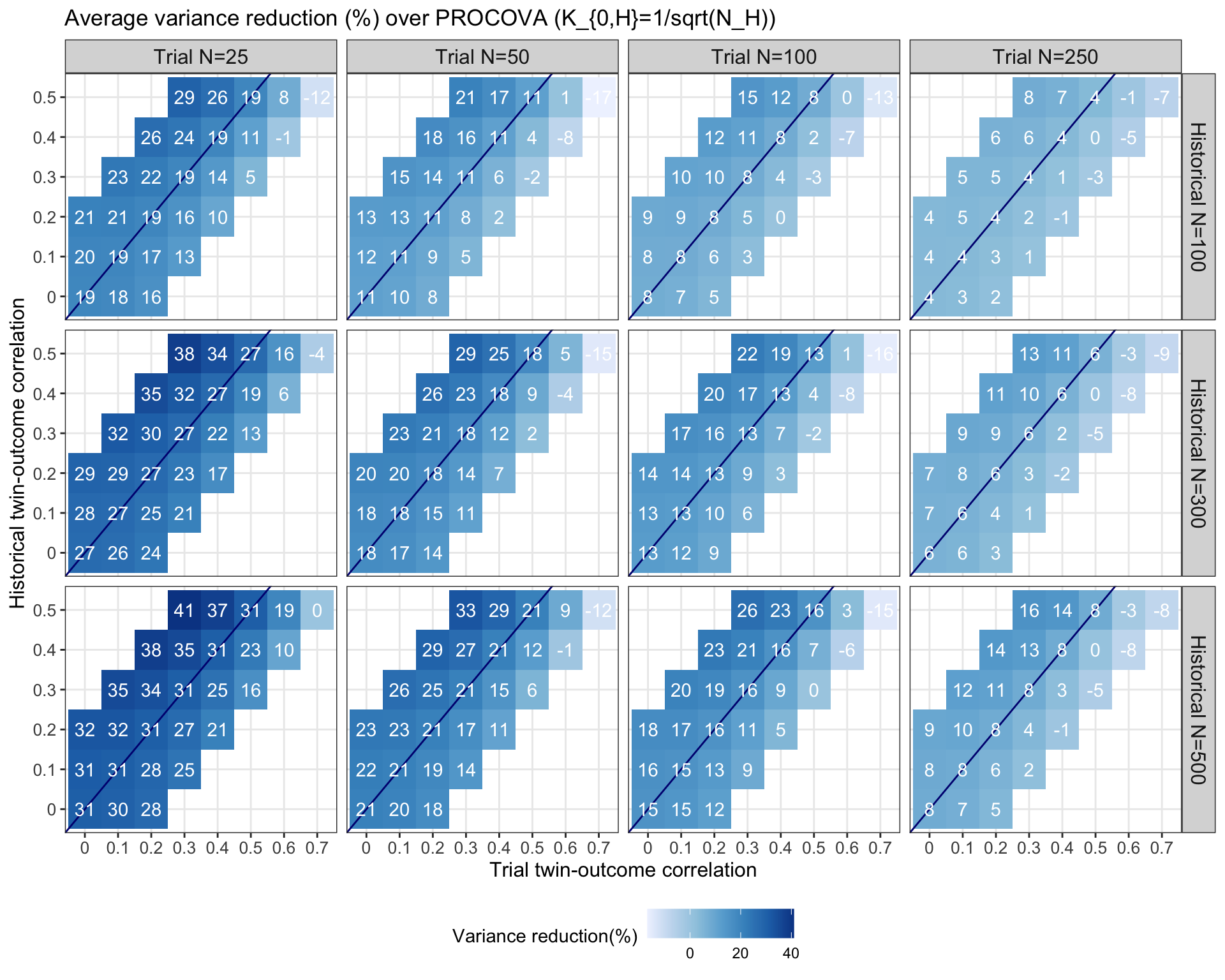}
\caption{Using $K_{0,H}=N_H^{-1/2}$: Average variance reduction (in percentages) of Bayesian PROCOVA over PROCOVA for different correlations of the prognostic scores with the control outcomes between historical and RCT data. The horizontal axis denotes the correlation in the RCT data and the vertical axis denotes the correlation in the historical data. The diagonal line in each panel represents cases in which the correlation is the same in both datasets.}
\label{fig:variance_reduction_correlation_shift-sqrt}
\end{figure}

Figures \ref{fig:variance_reduction_correlation_shift} and \ref{fig:variance_reduction_correlation_shift-sqrt} illustrate how the variance reduction of Bayesian PROCOVA changes due to discrepancies in the correlations between prognostic scores and control outcomes across the historical and RCT data. In general, combinations of small correlation levels (the bottom left sections of each panel) result in less variance reduction. Combinations of correlation levels that lie above the line, representing cases in which the correlation between prognostic scores and control outcomes in the historical data is larger than that in the RCT data, correspond to greater variance reduction. By comparing these two figures, we observe that the variance reduction resulting from $K_{0,H} = N_H^{-1/2}$ is less than that resulting from $K_{0,H} = N_H^{-1}$. This is a direct consequence of the fact that the prior distribution under the first setting of $K_{0,H}$ is less informative than that under the second setting, and hence the posterior for the treatment effect has greater posterior variance.

\begin{figure}[H]
\centering
\includegraphics[scale=0.2]{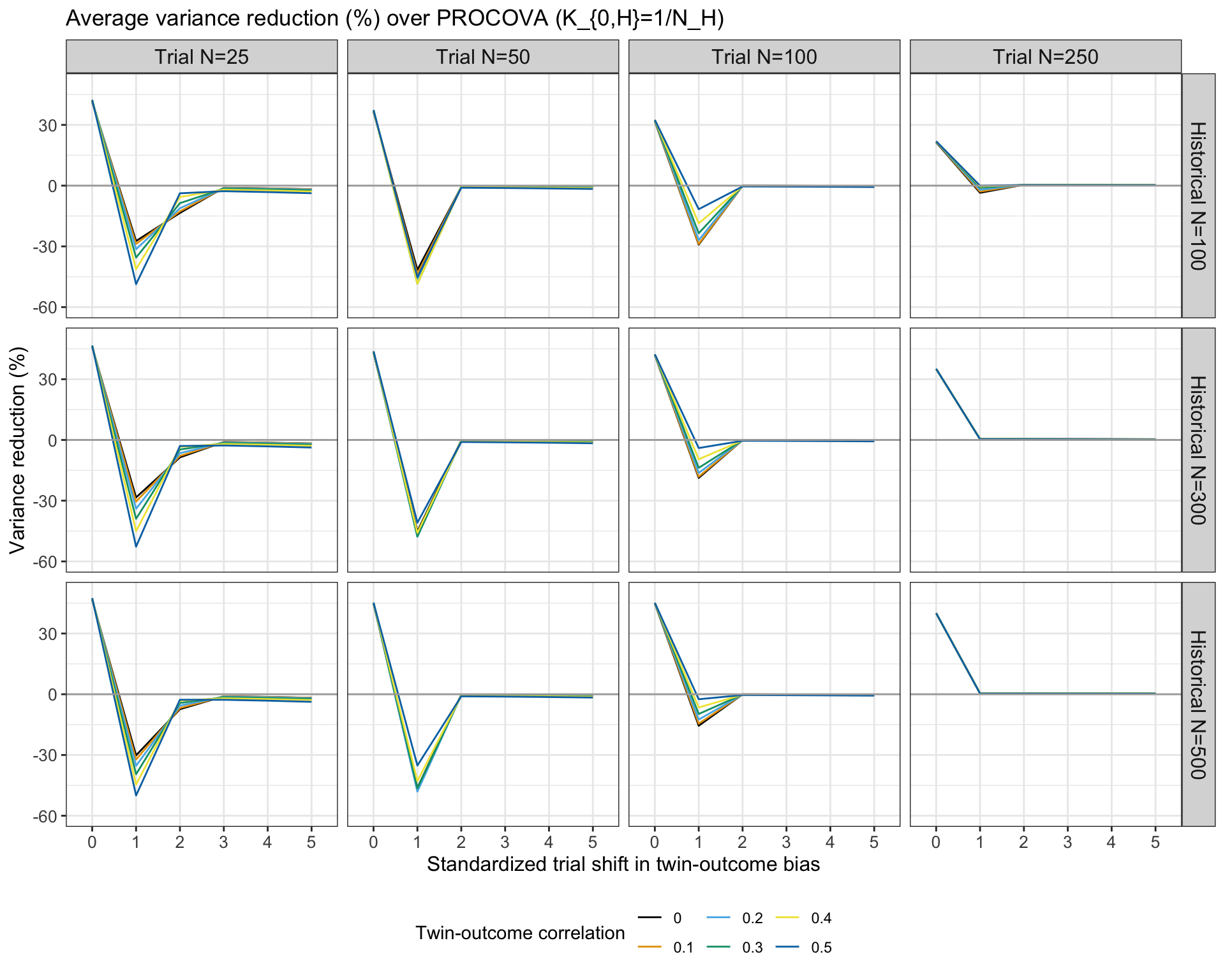}
\caption{Using $K_{0,H}=N_H^{-1}$: Average variance reduction (in percentages) of Bayesian PROCOVA over PROCOVA under a shift in the bias of the prognostic scores. In this case, the Inverse Chi-Square prior with $\nu_0 = 1, \sigma_0^2 = 1$ is specified for $\sigma^2$ in the flat prior component. The horizontal axis captures the absolute standardized shift in bias and the vertical axis indicates the percent variance reduction over PROCOVA. Each panel corresponds to a combination of RCT sample size (columns) and historical data sample size (rows). The line colors denote the correlation values between the prognostic scores and control outcomes, which are set to be the same in historical and RCT data.}
\label{fig:variance_reduction_bias_shift}
\end{figure}

\begin{figure}[H]
\centering
\includegraphics[scale=0.2]{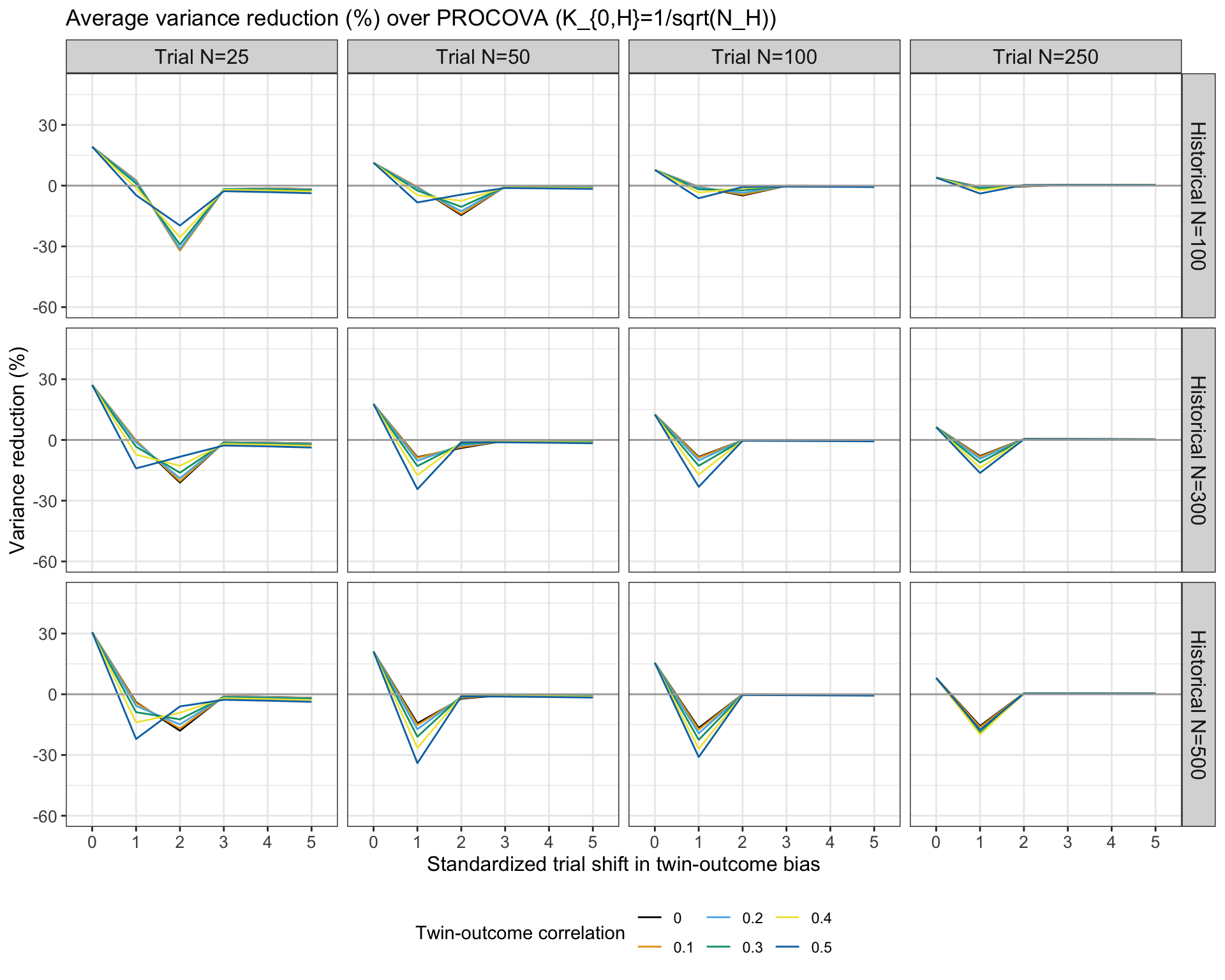}
\caption{Using $K_{0,H}=N_H^{-1/2}$: Average variance reduction (in percentages) of Bayesian PROCOVA over PROCOVA under a shift in the bias of the prognostic scores. In this case, the Inverse Chi-Square prior with $\nu_0 = 1, \sigma_0^2 = 1$ is specified for $\sigma^2$ in the flat prior component. The horizontal axis captures the absolute standardized shift in bias and the vertical axis indicates the percent variance reduction over PROCOVA. Each panel corresponds to a combination of RCT sample size (columns) and historical data sample size (rows). The line colors denote the correlation values between the prognostic scores and control outcomes, which are set to be the same in historical and RCT data.}
\label{fig:variance_reduction_bias_shift-sqrt}
\end{figure}

The case of a shift in the bias of the prognostic scores, i.e., a change in the intercept from the historical control to the RCT data, is demonstrated via Figures \ref{fig:variance_reduction_bias_shift} and \ref{fig:variance_reduction_bias_shift-sqrt}. We observe from these figures that when there’s no shift in the bias of the prognostic scores, Bayesian PROCOVA exhibits variance reduction over PROCOVA. However, in cases of a mild bias shift (e.g., one standardized unit), Bayesian PROCOVA can actually inflate the variance of the treatment effect estimator compared to PROCOVA. However, as the magnitude of the shift increases beyond one standardized unit, Bayesian PROCOVA effectively recovers the same inferences as PROCOVA, so that there would be zero variance reduction. Furthermore, when $K_{0,H} = N_H^{-1/2}$, the inflation of the variance of the treatment effect estimator is less than when $K_{0,H} = N_H^{-1}$. In addition, the severity and risk of variance inflation decreases as a function of the RCT sample size.

\begin{figure}[H]
\centering
\includegraphics[scale=0.2]{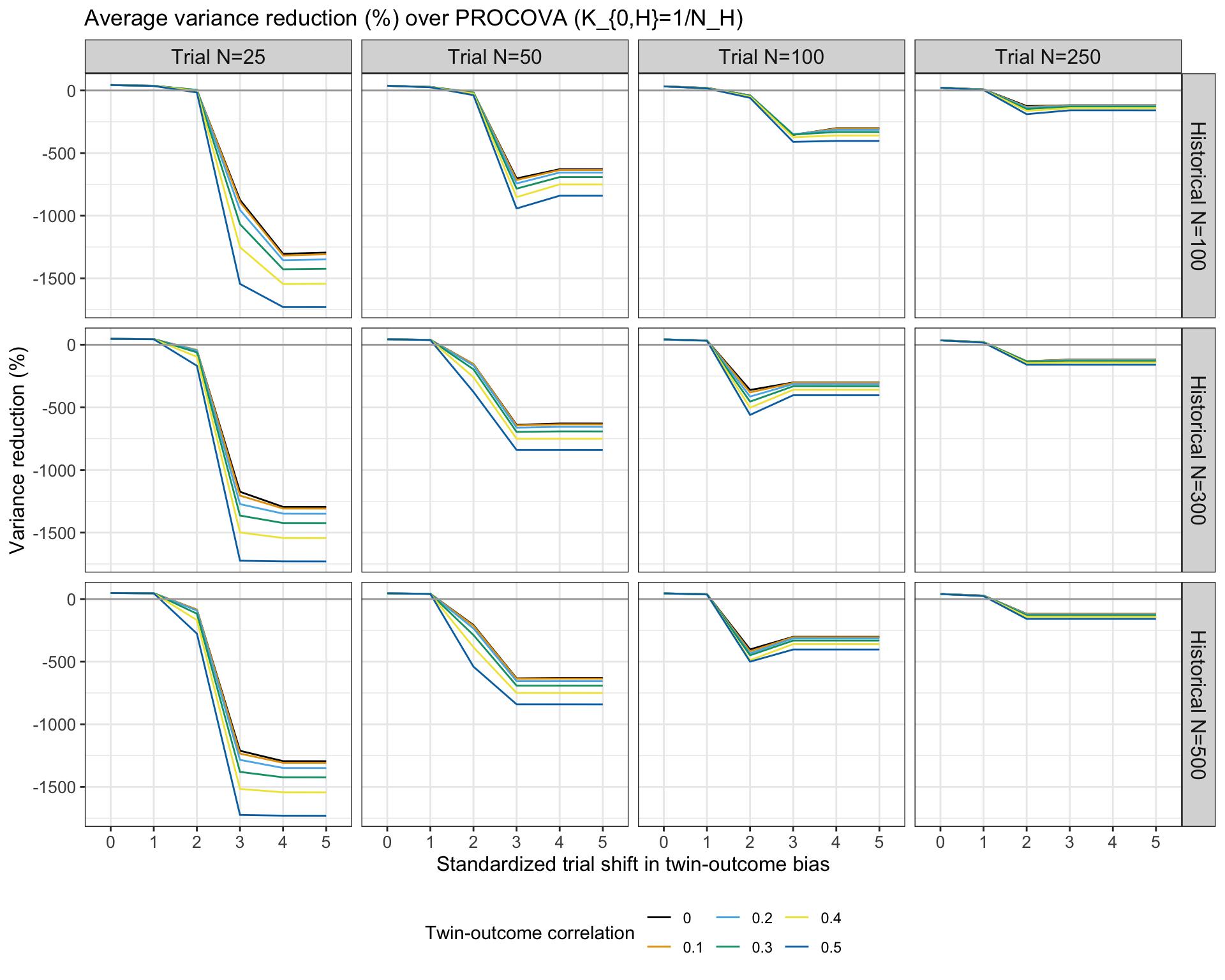}
\caption{Using $K_{0,H}=N_H^{-1}$: Average variance reduction (in percentages) of Bayesian PROCOVA over PROCOVA in the case of a shift in the prognostic score bias, and when the Inverse Chi-Square prior with $\nu_0 = 3, \sigma_0^2 = 100$ is placed on $\sigma^2$ in the flat prior component. The horizontal axis is the absolute standardized shift in prognostic score bias and the vertical axis in the percent of variance reduction for Bayesian PROCOVA compared to PROCOVA. The panels correspond to a combination of RCT sample size (rows) and historical control data sample size (columns). The line colors denote the correlations between the prognostic scores and the historical control data, which are set to be the same in both datasets.}
\label{fig:variance_reduction_bias_shift_invchisq3100}
\end{figure}

\begin{figure}[H]
\centering
\includegraphics[scale=0.2]{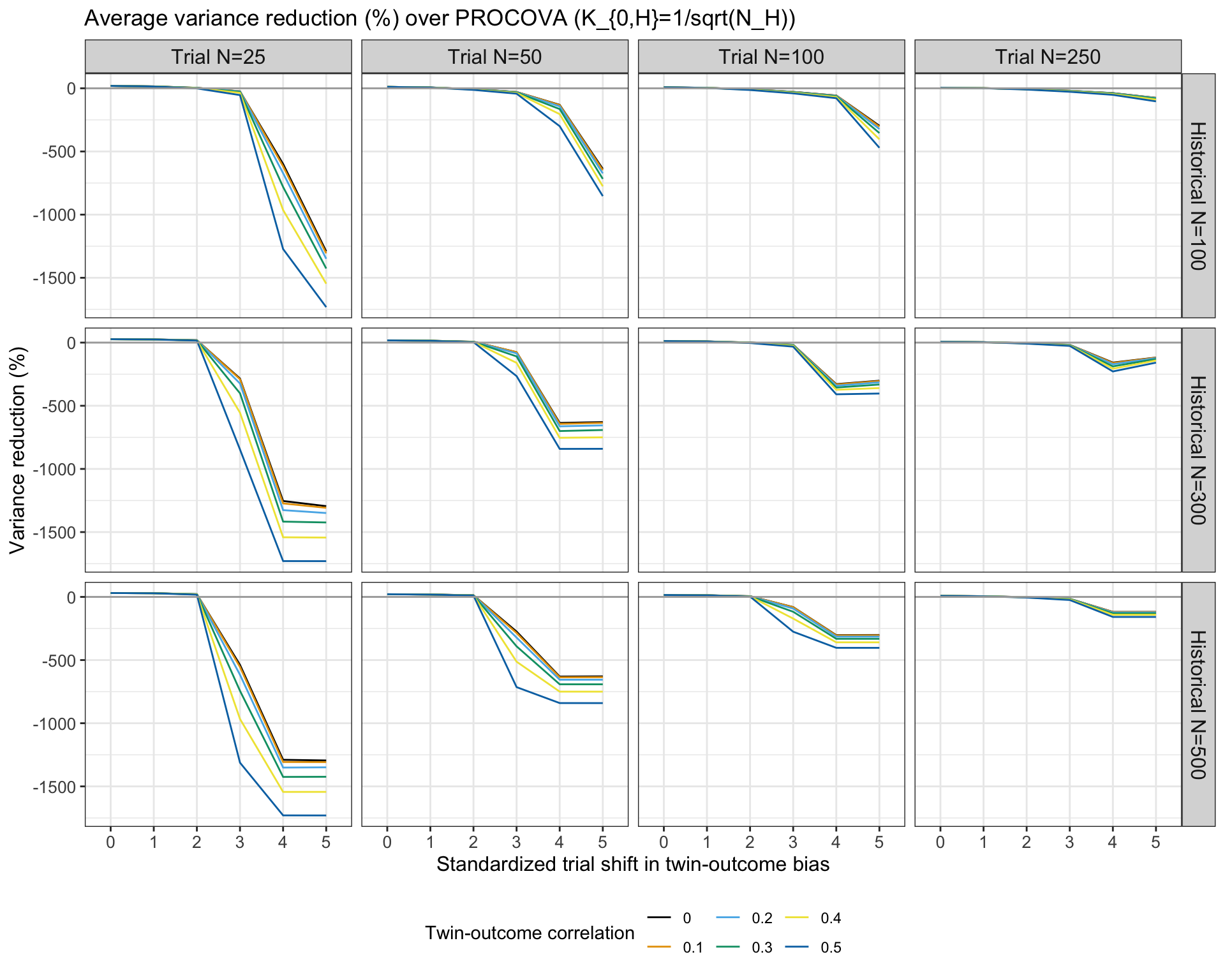}
\caption{Using $K_{0,H}=N_H^{-1/2}$: Average variance reduction (in percentages) of Bayesian PROCOVA over PROCOVA in the case of a shift in the prognostic score bias, and when the Inverse Chi-Square prior with $\nu_0 = 3, \sigma_0^2 = 100$ is placed on $\sigma^2$ in the flat prior component. The horizontal axis is the absolute standardized shift in prognostic score bias and the vertical axis in the percent of variance reduction for Bayesian PROCOVA compared to PROCOVA. The panels correspond to a combination of RCT sample size (rows) and historical control data sample size (columns). The line colors denote the correlations between the prognostic scores and the historical control data, which are set to be the same in both datasets.}
\label{fig:variance_reduction_bias_shift_invchisq3100-sqrt}
\end{figure}

The previous set of results indicate how one can mitigate variance inflation in cases of mild-to-moderate conflicts between historical control and RCT data by consideration of various hyperparameters in the prior specification for Bayesian PROCOVA. For example, in the previous simulation scenarios in which the prognostic score bias differed between the historical and RCT data, if we were to specify the Inverse Chi-Square prior with $\nu_0 = 3, \sigma_0^2 = 100$, then we would observe significant variance inflation of Bayesian PROCOVA over PROCOVA. This is illustrated in Figures \ref{fig:variance_reduction_bias_shift_invchisq3100} and \ref{fig:variance_reduction_bias_shift_invchisq3100-sqrt}. Ultimately, this choice of hyperparameters would lead to more weight being placed on the information from the historical control data, which decreases the quality of inferences due to the discrepancies between the historical control and RCT data. 

In contrast, by specifying the Inverse Chi-Square prior with $\nu_0 = 1, \sigma_0^2 = 1$, which more closely resembles the standard non-informative prior $p \left ( \sigma^2 \right ) \propto \left ( \sigma^2 \right )^{-1}$, we recover PROCOVA in cases of data conflict, as demonstrated in Figures \ref{fig:variance_reduction_bias_shift} and \ref{fig:variance_reduction_bias_shift-sqrt}. This indicates that we can quickly recover PROCOVA, even in cases of mild shift, by placing a tighter Inverse Chi-Square prior for $\sigma^2$ in the flat prior component. We can also interpret the tighter prior as a tighter penalty on the variance parameter, which prevents the shift in bias from entering the inferences for the variance term. 

We also observe from the Figures \ref{fig:variance_reduction_bias_shift_invchisq3100} and \ref{fig:variance_reduction_bias_shift_invchisq3100-sqrt} that there exists a potential trade-off with the stability of the variance reduction. Therefore the Inverse Chi-Square hyperparameter selection should be one focus of prospective sensitivity analyses. Another focus should be the selection of $K_{0,H}$, as changes in this hyperparameter can assist in discounting historical control data in case of domain shift.

\section{Concluding Remarks}
\label{sec:concluding_remarks}

The capability for effective and rapid decision-making from RCTs can be improved by means of innovative statistical methods that increase the precision of treatment effect inferences while controlling bias. Our Bayesian PROCOVA methodology directly addresses these crucial requirements for decision-making. It incorporates covariate adjustment based on optimized, generative AI algorithms under the Bayesian paradigm. The combination of these two strategies in Bayesian PROCOVA follows regulatory guidance and best practices on covariate adjustment and Bayesian inference. Key features of Bayesian PROCOVA are its additive mixture prior on the regression parameters, and the prior for the mixture weight. The complete prior specification encodes historical control information while also enabling the consideration of a weakly informative prior component in case discrepancies exist between the historical control and RCT data. The prior for the mixture weight is completely pre-specifiable prior to the commencement of the RCT, so that the RCT data is not used twice in Bayesian PROCOVA as in the methods of \citet{egidi_2022}, \citet{zhao_2023}, and \citet{yang_2023}. We derived the posterior distributions for the regression coefficients in closed-form conditional on the mixture weight, with treatment inferences formulated in terms of the super-population average treatment effect $\beta_1$. Our derivations led to the development of a straightforward and efficient Gibbs algorithm for sampling from the joint posterior distribution of all model parameters, including the mixture weight. Finally, we demonstrated via comprehensive simulation studies that Bayesian PROCOVA can be tuned to both control the bias and reduce the variance of its treatment effect inferences compared to PROCOVA. Ultimately, fewer control participants would be necessary for recruitment into the RCT, the RCT can consequently run much faster, and effective decision-making from RCTs can be accelerated by means of Bayesian PROCOVA.

\bibliographystyle{chicago}
\bibliography{bibliography}

\begin{thebibliography}{}

\bibitem[\protect\citeauthoryear{Cook, Gelman, and Rubin}{Cook
  et~al.}{2006}]{cook_2006}
Cook, S.~R., A.~Gelman, and D.~B. Rubin (2006).
\newblock Validation of software for {B}ayesian models using posterior
  quantiles.
\newblock {\em Journal of Computational and Graphical Statistics\/}~{\em
  15\/}(3), 675--692.

\bibitem[\protect\citeauthoryear{Egidi, Pauli, and Torelli}{Egidi
  et~al.}{2022}]{egidi_2022}
Egidi, L., F.~Pauli, and N.~Torelli (2022).
\newblock Avoiding prior–data conflict in regression models via mixture
  priors.
\newblock {\em Canadian Journal of Statistics\/}~{\em 50\/}(2), 491--510.

\bibitem[\protect\citeauthoryear{{European Medicines Agency}}{{European
  Medicines Agency}}{2015}]{ema_adjusting_2015}
{European Medicines Agency} (2015).
\newblock Guideline on {Adjustment} for {Baseline} {Covariates} in {Clinical}
  {Trials}.
\newblock
  https://www.ema.europa.eu/en/documents/scientific-guideline/guideline-adjustment-baseline-covariates-clinical-trials\_en.pdf.

\bibitem[\protect\citeauthoryear{{European Medicines Agency}}{{European
  Medicines Agency}}{2022}]{ema_procova_2022}
{European Medicines Agency} (2022).
\newblock Qualification {Opinion} for {Prognostic} {Covariate} {Adjustment}.
\newblock
  https://www.ema.europa.eu/en/documents/regulatory-procedural-guideline/qualification-opinion-prognostic-covariate-adjustment-procovatm\_en.pdf.

\bibitem[\protect\citeauthoryear{Fernando, Menon, Jansen, Naik, Nucci, Roberts,
  Wu, and Dolsten}{Fernando et~al.}{2022}]{fernando_2022}
Fernando, K., S.~Menon, K.~Jansen, P.~Naik, G.~Nucci, J.~Roberts, S.~S. Wu, and
  M.~Dolsten (2022).
\newblock Achieving end-to-end success in the clinic: Pfizer’s learnings on
  {R\&D} productivity.
\newblock {\em Drug Discovery Today\/}~{\em 27\/}(3), 697--704.

\bibitem[\protect\citeauthoryear{Fisher}{Fisher}{2023}]{fisher_2023}
Fisher, C.~K. (2023, May).
\newblock Medicine needs visionaries.
\newblock https://unlearnai.substack.com/p/medicine-needs-visionaries.

\bibitem[\protect\citeauthoryear{Fogel}{Fogel}{2018}]{fogel_2018}
Fogel, D. (2018).
\newblock Factors associated with clinical trials that fail and opportunities
  for improving the likelihood of success: A review.
\newblock {\em Contemp Clin Trials Commun.\/}~{\em 11}, 156--164.

\bibitem[\protect\citeauthoryear{{Food and Drug Administration}, {US Department
  of Health and Human Services}, {Center for Devices and Radiological Health
  (CDRH)}, and {Center for Biologics Evaluation and Research (CBER)}}{{Food and
  Drug Administration}
  et~al.}{2010}]{food_and_drug_administration_bayesian_2010}
{Food and Drug Administration}, {US Department of Health and Human Services},
  {Center for Devices and Radiological Health (CDRH)}, and {Center for
  Biologics Evaluation and Research (CBER)} (2010, February).
\newblock Guidance for the {U}se of {B}ayesian {S}tatistics in {M}edical
  {D}evice {C}linical {T}rials.
\newblock
  https://www.fda.gov/regulatory-information/search-fda-guidance-documents/guidance-use-bayesian-statistics-medical-device-clinical-trials.

\bibitem[\protect\citeauthoryear{{Food and Drug Administration}, {US Department
  of Health and Human Services}, {Center for Drug Evaluation and Research
  (CDER)}, and {Center for Biologics Evaluation and Research (CBER)}}{{Food and
  Drug Administration}
  et~al.}{2023}]{food_and_drug_administration_adjusting_2023}
{Food and Drug Administration}, {US Department of Health and Human Services},
  {Center for Drug Evaluation and Research (CDER)}, and {Center for Biologics
  Evaluation and Research (CBER)} (2023, May).
\newblock Adjusting for {Covariates} in {Randomized} {Clinical} {Trials} for
  {Drugs} and {Biological} {Products}: {Guidance} for {Industry}.
\newblock
  https://www.fda.gov/regulatory-information/search-fda-guidance-documents/adjusting-covariates-randomized-clinical-trials-drugs-and-biological-products.

\bibitem[\protect\citeauthoryear{Freedman}{Freedman}{1999}]{freedman_1999}
Freedman, D. (1999).
\newblock From association to causation: {S}ome remarks on the history of
  statistics.
\newblock {\em Statistical Science\/}~{\em 14\/}(3), 243--258.

\bibitem[\protect\citeauthoryear{Frieden}{Frieden}{2017}]{frieden_2017}
Frieden, T.~R. (2017).
\newblock Evidence for health decision making — beyond randomized, controlled
  trials.
\newblock {\em New England Journal of Medicine\/}~{\em 377\/}(5), 465--475.
\newblock PMID: 28767357.

\bibitem[\protect\citeauthoryear{Gelman, Carlin, Stern, Dunson, Vehtari, and
  Rubin}{Gelman et~al.}{2013}]{gelman_2013}
Gelman, A., J.~B. Carlin, H.~S. Stern, D.~B. Dunson, A.~Vehtari, and D.~B.
  Rubin (2013).
\newblock {\em Bayesian Data Analysis\/} (3 ed.).
\newblock Chapman and Hall/CRC.

\bibitem[\protect\citeauthoryear{Geman and Geman}{Geman and
  Geman}{1984}]{geman_1984}
Geman, S. and D.~Geman (1984).
\newblock Stochastic relaxation, gibbs distributions, and the bayesian
  restoration of images.
\newblock {\em IEEE Transactions on Pattern Analysis and Machine
  Intelligence\/}~{\em PAMI-6\/}(6), 721--741.

\bibitem[\protect\citeauthoryear{Han, Zhan, Zhong, Liu, and Lindborg}{Han
  et~al.}{2017}]{han_2017}
Han, B., J.~Zhan, Z.~J. Zhong, D.~L. Liu, and S.~Lindborg (2017).
\newblock Covariate-adjusted borrowing of historical control data in randomized
  clinical trials.
\newblock {\em Pharmaceutical Statistics\/}~{\em 16\/}(4), 296--308.

\bibitem[\protect\citeauthoryear{Hobbs, Carlin, and Sargent}{Hobbs
  et~al.}{2013}]{hobbs_2013}
Hobbs, B., B.~Carlin, and D.~Sargent (2013).
\newblock Adaptive adjustment of the randomization ratio using historical
  control data.
\newblock {\em Clinical Trials\/}~{\em 10\/}(3), 430--440.

\bibitem[\protect\citeauthoryear{Holland}{Holland}{1986}]{holland_1986}
Holland, P.~W. (1986).
\newblock Statistics and causal inference.
\newblock {\em Journal of the American Statistical Association\/}~{\em
  81\/}(396), 945--960.

\bibitem[\protect\citeauthoryear{Imbens and Rubin}{Imbens and
  Rubin}{2015}]{imbens_rubin_2015}
Imbens, G.~W. and D.~B. Rubin (2015).
\newblock {\em Causal Inference for Statistics, Social, and Biomedical
  Sciences: An Introduction\/} (1 ed.).
\newblock Cambridge University Press.

\bibitem[\protect\citeauthoryear{Ionan, Clark, Travis, Amatya, Scott, Smith,
  Chattopadhyay, Salerno, and Rothmann}{Ionan et~al.}{2023}]{ionan_2023}
Ionan, A.~C., J.~Clark, J.~Travis, A.~Amatya, J.~Scott, J.~P. Smith,
  S.~Chattopadhyay, M.~J. Salerno, and M.~Rothmann (2023).
\newblock Bayesian methods in human drug and biological products development in
  {CDER} and {CBER}.
\newblock {\em Therapeutic Innovation \& Regulatory Science\/}~{\em 57\/}(3),
  436--444.

\bibitem[\protect\citeauthoryear{Kaizer, Koopmeiners, and Hobbs}{Kaizer
  et~al.}{2018}]{kaizer_2018}
Kaizer, A., J.~Koopmeiners, and B.~Hobbs (2018).
\newblock Bayesian hierarchical modeling based on multisource exchangeability.
\newblock {\em Biostatistics\/}~{\em 19\/}(2), 169--184.

\bibitem[\protect\citeauthoryear{Muehlemann, Zhou, Mukherjee, Hossain,
  Roychoudhury, and Russek-Cohen}{Muehlemann et~al.}{2023}]{muehlemann_2023}
Muehlemann, N., T.~Zhou, R.~Mukherjee, M.~I. Hossain, S.~Roychoudhury, and
  E.~Russek-Cohen (2023).
\newblock A tutorial on modern {B}ayesian methods in clinical trials.
\newblock {\em Therapeutic Innovation \& Regulatory Science\/}~{\em 57\/}(3),
  402--416.

\bibitem[\protect\citeauthoryear{Polack, Thomas, Kitchin, Absalon, Gurtman,
  Lockhart, Perez, P\'{e}rez~Marc, Moreira, Zerbini, Bailey, Swanson,
  Roychoudhury, Koury, Li, Kalina, Cooper, Frenck, Hammitt, T\"{u}reci, Nell,
  Schaefer, \"{U}nal, Tresnan, Mather, Dormitzer, \c{S}ahin, Jansen, and
  Gruber}{Polack et~al.}{2020}]{polack_2020}
Polack, F.~P., S.~J. Thomas, N.~Kitchin, J.~Absalon, A.~Gurtman, S.~Lockhart,
  J.~L. Perez, G.~P\'{e}rez~Marc, E.~D. Moreira, C.~Zerbini, R.~Bailey, K.~A.
  Swanson, S.~Roychoudhury, K.~Koury, P.~Li, W.~V. Kalina, D.~Cooper, R.~W.
  Frenck, L.~L. Hammitt, O.~T\"{u}reci, H.~Nell, A.~Schaefer, S.~\"{U}nal,
  D.~B. Tresnan, S.~Mather, P.~R. Dormitzer, U.~\c{S}ahin, K.~U. Jansen, and
  W.~C. Gruber (2020).
\newblock Safety and efficacy of the {BNT162b2} m{RNA} {C}ovid-19 vaccine.
\newblock {\em New England Journal of Medicine\/}~{\em 383\/}(27), 2603--2615.
\newblock PMID: 33301246.

\bibitem[\protect\citeauthoryear{Romano and Wolf}{Romano and
  Wolf}{2017}]{romano_2017}
Romano, J.~P. and M.~Wolf (2017).
\newblock Resurrecting weighted least squares.
\newblock {\em Journal of Econometrics\/}~{\em 197\/}(1), 1--19.

\bibitem[\protect\citeauthoryear{Rosenbaum and Rubin}{Rosenbaum and
  Rubin}{1983}]{rosenbaum_1983}
Rosenbaum, P. and D.~B. Rubin (1983).
\newblock The central role of the propensity score in observational studies for
  causal effects.
\newblock {\em Biometrika\/}~{\em 70}, 41--55.

\bibitem[\protect\citeauthoryear{Rubin}{Rubin}{1974}]{rubin_1974}
Rubin, D.~B. (1974).
\newblock Estimating causal effects of treatments in randomized and
  nonrandomized studies.
\newblock {\em Journal of Educational Psychology\/}~{\em 66\/}(5), 688--701.

\bibitem[\protect\citeauthoryear{Rubin}{Rubin}{1978}]{rubin_1978}
Rubin, D.~B. (1978).
\newblock {Bayesian Inference for Causal Effects: The Role of Randomization}.
\newblock {\em The Annals of Statistics\/}~{\em 6\/}(1), 34--58.

\bibitem[\protect\citeauthoryear{Schuler, Walsh, Hall, Walsh, Fisher,
  Initiative, et~al.}{Schuler et~al.}{2022}]{schuler_2022}
Schuler, A., D.~Walsh, D.~Hall, J.~Walsh, C.~Fisher, A.~D.~N. Initiative,
  et~al. (2022).
\newblock Increasing the efficiency of randomized trial estimates via linear
  adjustment for a prognostic score.
\newblock {\em The International Journal of Biostatistics\/}~{\em 18\/}(2),
  329--356.

\bibitem[\protect\citeauthoryear{Splawa-Neyman, Dabrowska, and
  Speed}{Splawa-Neyman et~al.}{1990}]{neyman_1923}
Splawa-Neyman, J., D.~M. Dabrowska, and T.~P. Speed (1990).
\newblock {On the Application of Probability Theory to Agricultural
  Experiments. Essay on Principles. Section 9}.
\newblock {\em Statistical Science\/}~{\em 5\/}(4), 465 -- 472.

\bibitem[\protect\citeauthoryear{Subbiah}{Subbiah}{2023}]{subbiah_2023}
Subbiah, V. (2023).
\newblock The next generation of evidence-based medicine.
\newblock {\em Nature Medicine\/}~{\em 29}, 49--58.

\bibitem[\protect\citeauthoryear{Travis, Rothmann, and Thomson}{Travis
  et~al.}{2023}]{travis_2023}
Travis, J., M.~Rothmann, and A.~Thomson (2023).
\newblock Perspectives on informative {B}ayesian methods in pediatrics.
\newblock {\em Journal of Biopharmaceutal Statistics\/}~{\em 29}, 1--14.

\bibitem[\protect\citeauthoryear{Tsiatis, Davidian, Zhang, and Lu}{Tsiatis
  et~al.}{2008}]{tsiatis_2008}
Tsiatis, A., M.~Davidian, M.~Zhang, and X.~Lu (2008).
\newblock Covariate adjustment for two-sample treatment comparisons in
  randomized trials: {A} principled yet flexible approach.
\newblock {\em Statistics in Medicine\/}~{\em 27\/}(23), 4658--4677.

\bibitem[\protect\citeauthoryear{Walsh, Schuler, Hall, Walsh, and Fisher}{Walsh
  et~al.}{2020}]{walsh_2020}
Walsh, D., A.~Schuler, D.~Hall, J.~Walsh, and C.~Fisher (2020).
\newblock Bayesian prognostic covariate adjustment.
\newblock https://arxiv.org/abs/2012.13112.

\bibitem[\protect\citeauthoryear{White}{White}{1980}]{white_1980}
White, H. (1980).
\newblock A heteroskedasticity-consistent covariance matrix estimator and a
  direct test for heteroskedasticity.
\newblock {\em Econometrica\/}~{\em 48\/}(4), 817--838.

\bibitem[\protect\citeauthoryear{Yang, Zhao, Nie, Vallejo, and Yuan}{Yang
  et~al.}{2023}]{yang_2023}
Yang, P., Y.~Zhao, L.~Nie, J.~Vallejo, and Y.~Yuan (2023).
\newblock {SAM}: {S}elf-adapting mixture prior to dynamically borrow
  information from historical data in clinical trials.
\newblock {\em Biometrics\/}, 1--32.

\bibitem[\protect\citeauthoryear{Yule}{Yule}{1899}]{yule_1899}
Yule, G.~U. (1899).
\newblock An investigation into the causes of changes in pauperism in
  {E}ngland, chiefly during the last two intercensal decades.
\newblock {\em Journal of the Royal Statistical Society\/}~{\em 62}, 249--295.

\bibitem[\protect\citeauthoryear{Zhao and Ma}{Zhao and Ma}{2023}]{zhao_2023}
Zhao, Q. and H.~Ma (2023).
\newblock Modified robust meta-analytic-predictive priors for incorporating
  historical controls in clinical trials.
\newblock {\em Statistics in Biopharmaceutical Research\/}, 1--7.

\end{thebibliography}

\newpage

\appendix

\section{Mathematical Details}
The probability density function for the informative prior component is 
\begin{align*}
p_I \left ( \beta, \sigma^2 \right ) &= \left \{ \frac{ \left ( \frac{N_H - 2}{2}\right )^{\left ( N_H-2 \right )/2} \left ( s_H^2 \right )^{\left ( N_H-2 \right )/2} \left ( \sigma^2 \right )^{-\left \{ (N_H+1)/2 + 1 \right \}}}{\Gamma \left ( \frac{N_H-2}{2} \right ) \pi^{3/2}  \left (K_{0,H} K_{1,H} K_{2,H} \right )^{1/2}} \right \}  \\
& \ \ \ \times \mathrm{exp} \left [ -\frac{(N_H-2)s_H^2}{2\sigma^2} - \frac{1}{2\sigma^2} \left \{ \frac{ \left ( \beta_0 - \widehat{\beta_{0,H}} \right )^2}{K_{0,H}} + \frac{\beta_1^2}{K_{1,H}} + \frac{ \left ( \beta_2 - \widehat{\beta_{2,H}} \right )^2}{K_{2,H}} \right \} \right ].
\end{align*}
The probability density function for the flat prior component is
\begin{align*}
    p_F \left ( \beta, \sigma^2 \right ) = \left \{ \frac{ \left ( \frac{\nu_0}{2} \right )^{\nu_0/2} \left ( \sigma_0^2 \right )^{\nu_0/2} \left ( \sigma^2 \right )^{-\left \{ (\nu_0 + 3)/2 +1 \right \}}}{\Gamma \left ( \frac{\nu_0}{2} \right ) \left ( \pi k \right)^{3/2}} \right \} \mathrm{exp} \left [ \frac{-\nu_0\sigma_0^2}{2\sigma^2} - \frac{1}{2k\sigma^2} \left ( \beta_0^2 + \beta_1^2 + \beta_2^2  \right ) \right ].
\end{align*}
The closed-form expression for the inverse of the normalization constant of the conditional posterior distribution of the regression parameters, $C^{-1}$, is
\begin{align*}
C^{-1} &= \left \{ \frac{\boldsymbol{\omega} \Gamma \left ( \frac{N+N_H-2}{2} \right )}{\Gamma \left ( \frac{N_H-2}{2} \right ) \left ( N_H-2 \right )^{N/2}\pi^{N/2}} \right \} \left ( s_H^2 \right )^{-N/2} \left \{ \mathrm{det} \left ( I_{N \times N} + V \boldsymbol{K}V^{\mathsf{T}} \right ) \right \}^{-1/2} \\
& \ \ \ \ \ \ \times \Bigg{\{} 1 + \frac{1}{\left ( N_H-2 \right ) s_H^2} \left ( y^{(c)} - V\begin{pmatrix} \widehat{\beta_{0,H}} \\ 0 \\ \widehat{\beta_{2,H}} \end{pmatrix} \right )^{\mathsf{T}}  \left ( I_{N \times N} + V \boldsymbol{K} V^{\mathsf{T}} \right )^{-1} \\
& \ \ \ \ \ \ \ \ \ \ \ \ \ \ \ \ \ \ \ \ \ \ \ \ \ \ \ \ \ \ \ \ \ \ \left ( y^{(c)} - V \begin{pmatrix} \widehat{\beta_{0,H}} \\ 0 \\ \widehat{\beta_{2,H}} \end{pmatrix} \right ) \Bigg{\}}^{-\left \{ \left ( N+N_H-2 \right )/2 \right \}} \\
& \ \ \ + \left \{ \frac{\left ( 1 - \boldsymbol{\omega} \right ) \Gamma \left ( \frac{N+\nu_0}{2} \right )}{\Gamma \left ( \frac{\nu_0}{2} \right ) \left ( \nu_0 \right )^{N/2}\pi^{N/2}} \right \} \left ( \sigma_0^2 \right )^{-N/2} \left \{ \mathrm{det} \left ( I_{N \times N} + kVV^{\mathsf{T}} \right ) \right \}^{-1/2} \\
& \ \ \ \ \ \ \times \left \{ 1 + \left ( \frac{1}{\nu_0 \sigma_0^2} \right ) \left ( y^{(c)} \right )^{\mathsf{T}}  \left ( I_{N \times N} + kVV^{\mathsf{T}} \right )^{-1} y^{(c)} \right \}^{-\left \{ \left ( N+\nu_0 \right )/2 \right \} }.
\end{align*}
The closed-form expression for the weight of the informative component in this conditional posterior distribution is
\begin{align*}
\boldsymbol{\omega}_* &= \left \{ \frac{C \boldsymbol{\omega} \Gamma \left ( \frac{N+N_H-2}{2} \right )}{\Gamma \left ( \frac{N_H-2}{2} \right ) \left ( N_H-2 \right )^{N/2}\pi^{N/2}} \right \} \left ( s_H^2 \right )^{-N/2} \left \{ \mathrm{det} \left ( I_{N \times N} + V \boldsymbol{K}V^{\mathsf{T}} \right ) \right \}^{-1/2} \\
& \ \ \ \times \Bigg{\{} 1 + \frac{1}{\left ( N_H-2 \right ) s_H^2} \left ( y^{(c)} - V \begin{pmatrix} \widehat{\beta_{0,H}} \\ 0 \\ \widehat{\beta_{2,H}} \end{pmatrix} \right )^{\mathsf{T}}  \left ( I_{N \times N} + V \boldsymbol{K} V^{\mathsf{T}} \right )^{-1} \\
& \ \ \ \ \ \ \ \ \ \ \ \ \ \ \ \ \ \ \ \ \ \ \ \ \ \ \ \ \ \ \ \left ( y^{(c)} - V \begin{pmatrix} \widehat{\beta_{0,H}} \\ 0 \\ \widehat{\beta_{2,H}} \end{pmatrix} \right ) \Bigg{\}}^{-\left \{ \left ( N+N_H-2 \right )/2 \right \} }
\end{align*}
and the closed-form expression for the weight of the flat component is
\begin{align*}
1-\boldsymbol{\omega}_* &= \left \{ \frac{C \left ( 1 - \boldsymbol{\omega} \right ) \Gamma \left ( \frac{N+\nu_0}{2} \right )}{\Gamma \left ( \frac{\nu_0}{2} \right ) \left ( \nu_0 \right )^{N/2}\pi^{N/2}} \right \} \left ( \sigma_0^2 \right )^{-N/2} \left \{ \mathrm{det} \left ( I_{N \times N} + kVV^{\mathsf{T}} \right ) \right \}^{-1/2} \\
& \ \ \ \times \left \{ 1 + \left ( \frac{1}{\nu_0 \sigma_0^2} \right ) \left ( y^{(c)} \right )^{\mathsf{T}}  \left ( I_{N \times N} + kVV^{\mathsf{T}} \right )^{-1} y^{(c)} \right \}^{-\left \{ \left ( N+\nu_0 \right )/2 \right \}}.
\end{align*}

\section{Specifying the Hyperparameters to Control the Operating Characteristics of Bayesian PROCOVA} 
\label{appendix:v2}
In its guidance on Bayesian methods for device trials, the FDA states that Type I error rates for Bayesian designs may exceed the typical thresholds adopted in frequentist analyses \citep[p.~29, 41]{food_and_drug_administration_bayesian_2010}. However, they also recommend that sponsors take steps to reduce or control Type I error rates so as to prevent them from being too large. One suggested step is to tune the prior so as to discount historical information when the amount of such information is disproportionately large compared to the information provided in trial \citep[p.~41]{food_and_drug_administration_bayesian_2010}. 

We recommend implementing this regulatory guidance in Bayesian PROCOVA via the specification of the hyperparameter values $K_{0,H}$ and $K_{2,H}$ in the informative component of the mixture prior, and keeping $\sigma^2 \sim (N_H-2)s_H^2/\chi_{N_H-2}^2$ as before. In particular, the value for $K_{0,H}$ directly affects the operating characteristics of Bayesian PROCOVA. Specifically, as the absolute magnitude of the twin-outcome bias between the historical control and RCT data increases, the rate and magnitude of the Type I error rate inflation, treatment effect estimate bias accumulation, and variance reduction over PROCOVA changes as well. These effects of $K_{0,H}$ on the operating characteristics are evident from the figures summarizing the results of the simulation studies in Section \ref{sec:simulation_experiments}. In general, if the baseline characteristics in the historical control data are well-matched to those in the prospective trial, then the bias of the twins in both datasets should be approximately equal because the DTG is effectively predicting the control outcomes for two datasets from the same population. This can hold on average for large sample sizes, but in small samples bias may arise due to the natural variation involved in finite historical and trial data samples.Therefore, in order to implement the regulatory guidance, a key task of Bayesian PROCOVA is to specify $K_{0,H}$ so as to account for potential twin-outcome bias shifts. 

To formally describe how $K_{0,H}$ and $K_{2,H}$ can be specified to implement regulatory guidance for Bayesian PROCOVA, let the twin-outcome bias shift be denoted by $\Delta = \beta_{0,T} - \beta_{0,H}$, where $\beta_{0,T}$ denotes the intercept parameter of the regression model underlying for the RCT and $\beta_{0,H}$ denotes the intercept parameter of the regression model underlying the historical control data. Assume that $\Delta \sim \mathrm{N} \left ( 0, \mathrm{Var} \left ( \Delta \right ) \right )$. In addition to the twin-outcome bias shift, we will introduce a more general scaling constant $\gamma$ to discount the historical control sample size for both $K_{0,H}$ and $K_{2,H}$. For example, in the simulation studies in Section \ref{sec:simulation_experiments}, we set $\gamma = N_H^{-1/2}$ for specifying $K_{0,H}$, but did not introduce such a scaling constant for $K_{2,H}$. We combine the propagation of the uncertainty regarding the twin-outcome bias shift to the RCT along with $\gamma$ to formulate the general hyperparameter specification as
\begin{align*}
K_{0,H} &= \frac{\gamma}{N_H} + \frac{\mathrm{Var}(\Delta)}{s^2_H}, \\
K_{2,H} &= \frac{\gamma}{N_H\displaystyle \sum_{i=1}^N \left ( m_{i,H} - \bar{m}_H \right )^2},
\end{align*}
where $s^2_H$ is defined in Section \ref{sec:Bayesian_PROCOVA_model} and $\gamma$ is a constant that scales the historical sample size $N_H$. The term $\mathrm{Var}(\Delta)/s^2_H$ scales the twin-outcome bias shift variance by the residual variance in the historical control data. This term can be determined by consideration of the respective sample sizes of the historical control and RCT data, and can be estimated through a bootstrap procedure on the historical control data. For example, one can sample the historical data with replacement at a sample size equal to that of the prospective RCT and obtain the value of $\hat{\beta}_{0,\text{boot}}$ from the linear regression $y_{i,\text{boot}} = \beta_{0,\text{boot}} + \beta_{2,\text{boot}}(m_i - \bar{m}) + \bar{m} + \epsilon_i$. Repeat this $J$ times and take $\mathrm{Var}(\Delta)$ as the variance of $\Tilde{\beta}_{\text{boot}} = \{\hat{\beta}_{0,\text{boot}}^{(1)},...,\hat{\beta}_{0,\text{boot}}^{(J)}\}$.

Given the selection of $\mathrm{Var}(\Delta)$ and $s_H^2$, $\gamma$ can be selected to control the expected operating characteristics of Bayesian PROCOVA. For example, the expected Type I error rate and variance reduction of Bayesian PROCOVA over PROCOVA can be estimated via simulation or bootstrap sampling of the historical data. It's critical to consider different values of twin-outcome bias shift in the hyperparameter specification, because for a given amount of twin-outcome bias shift a value for $\gamma$ can be chosen to yield a desired Type I error rate and corresponding expected variance reduction of Bayesian PROCOVA over PROCOVA. Given the assumption that $\Delta \sim \mathrm{N} \left ( 0, \mathrm{Var} \left ( \Delta \right ) \right )$, one can define a maximum twin-outcome bias shift with respect to a certain number of standard deviations. The advantage of this consideration is that it would enable one to specify $K_{0,H}$ such that, for example, the maximum Type I error rate is controlled at $10\%$ for twin-outcome bias shifts up to four standard deviations of $\Delta$. As demonstrated in the results of the simulations from Section \ref{sec:simulation_experiments}, larger values of $K_{0,H}$ yield more conservative performance in the form of smaller Type I error rate and more modest variance reduction. The specification of $K_{2,H}$ involves the scaling factor of $\frac{\gamma}{N_H}$ for consistency.

\end{document}